\documentclass[a4paper,11pt]{article}
\pdfoutput=1 
\usepackage{jcappub}

\usepackage{natbib}
\usepackage[T1]{fontenc} 
\usepackage{amsmath}
\usepackage{graphicx}
\usepackage[utf8]{inputenc}
\usepackage{hyperref}
\hypersetup{colorlinks=true,linkcolor=blue,citecolor=magenta,filecolor=magenta, urlcolor=cyan}
\usepackage{amsmath}
\usepackage{graphicx}
\RequirePackage{color}
\usepackage{cleveref}

\title{Robust Limits from Upcoming Neutrino Telescopes and Implications on Minimal Dark Matter Models}

\author[a]{S.~Basegmez du Pree,}
\author[b]{C.~Arina,}
\author[b]{A.~Cheek,}
\author[c]{A.~Dekker,}
\author[c,d,e]{M.~Chianese}
\author[c,f]{and S.~Ando}

\date{\today}

\affiliation[a]{Nikhef, National Institute for Subatomic Physics, Science Park 105
1098 XG, Amsterdam, The Netherlands}
\affiliation[b]{Centre for Cosmology, Particle Physics and Phenomenology (CP3), Universit\'e catholique de Louvain, Chemin du Cyclotron 2, B-1348 Louvain-la-Neuve, Belgium}
\affiliation[c]{GRAPPA Institute, University of Amsterdam, 1098 XH Amsterdam, The Netherlands}
\affiliation[d]{Dipartimento di Fisica "Ettore Pancini", Università degli studi di Napoli "Federico II", Complesso Univ. Monte S. Angelo, I-80126 Napoli, Italy}
\affiliation[e]{INFN - Sezione di Napoli, Complesso Univ. Monte S. Angelo, I-80126 Napoli, Italy}
\affiliation[f]{Kavli Institute for the Physics and Mathematics of the Universe (Kavli IPMU, WPI), University of Tokyo, Kashiwa, Chiba 277-8583, Japan}

\emailAdd{s.basegmez.du.pree@nikhef.nl}
\emailAdd{chiara.arina@uclouvain.be}
\emailAdd{andrew.cheek@uclouvain.be}
\emailAdd{a.h.dekker@uva.nl}
\emailAdd{marco.chianese@unina.it}

\newcommand{\sigmav}{\langle \sigma v \rangle}
\newcommand{\ie}{{\it i.e.}\ }
\newcommand{\eg}{{\it e.g.}\ }

\abstract{
Experimental developments in neutrino telescopes are drastically improving their ability to constrain the annihilation cross-section of dark matter. In this paper, we employ an angular power spectrum analysis method to probe the galactic and extra-galactic dark matter signals with neutrino telescopes. We first derive projections for a next generation of neutrino telescope that is inspired by KM3NeT. We emphasise that such analysis is much less sensitive to the choice of dark matter density profile. 
Remarkably, the projected sensitivity is improved by more than an order of magnitude with respect to the existing limits obtained by assuming the Burkert dark matter density profile describing the galactic halo. 
Second, we analyse minimal extensions to the Standard Model that will be maximally probed by the next generation of neutrino telescopes. As benchmark scenarios, we consider Dirac dark matter in $s$- and $t$-channel models with vector and scalar mediators. We follow a global approach by examining all relevant complementary experimental constraints.
We find that neutrino telescopes will be able to competitively probe significant portions of parameter space. Interestingly, the anomaly-free $L_{\mu}-L_{\tau}$ model can potentially be explored in regions where the relic abundance is achieved through freeze-out mechanism. 
}

\keywords{Dark Matter, Neutrinos, Neutrino Telescopes, Minimal Models}

\begin{document}

\maketitle

\section{Introduction}\label{sec:intro}

The true nature of dark matter is one of the prevailing problems in modern physics. Searches for signatures of the dark matter interacting with the Standard Model (SM) particles, so far being empty handed by direct and collider experiments, suggest us to think perhaps dark matter can have different interactions that have not been in our reach. Although, thermal production mechanisms provide a good motivation for testing non-gravitational interactions between dark matter and the SM particles, yet the experimental landscape providing only the exclusion limits encourages one to assess the possibilities that remain. 

Despite neutrinos being a primary astrophysical messengers, their sensitivity on dark matter annihilation cross-sections is often overlooked. This is in large part due to detector sensitivities, see \eg ref.~\cite{Arguelles:2019ouk} for a recent review. Although current neutrino experiments are not sensitive enough to probe a thermally produced dark matter annihilation signal in the context of WIMPs (Weakly Interacting Massive Particles), experimental efforts are drastically altering the situation. The next generation of neutrino telescopes under construction will change this picture. Although this work can be extended to different types of neutrino experiments, such as IceCube Upgrade~\cite{Baur:2019jwm}, IceCube-Gen 2~\cite{Aartsen:2019swn}, Baikal-GVD~\cite{Avrorin:2014vca}, P-ONE~\cite{Agostini:2020aar} and KM3NeT~\cite{Adri_n_Mart_nez_2016}, as a matter of choice we focus on the potential of the latter neutrino telescope. KM3NeT is a neutrino telescope which comprises a cubic kilometer volume of optical arrays, currently being constructed in the deep Mediterranean Sea and already partially operating. This choice is particularly interesting as KM3NeT not only has a superior angular resolution but also a good field of view to the galactic centre, which is assumed to be one of the most promising target for dark matter searches because of the large dark matter density.

The first part of this paper develops on the idea that dark matter annihilation can give observable cosmic neutrino signals which can be verified by different analysis techniques. 
Standard indirect dark matter searches, which examine small regions around the galactic center, typically suffer from large systematic uncertainties due to our little understanding of the dark matter halo in the inner regions of the Milky Way~\cite{Benito:2019ngh,Benito:2020lgu}. Depending on the choice of the dark matter profile, the limits on the dark matter annihilation cross-section can indeed differ by a few orders of magnitude~\cite{Abbasi:2011eq,Adrian-Martinez:2015wey,Aartsen:2017ulx,Albert:2016emp,ANTARES:2019svn,Aartsen:2020tdl}. This prevents one in setting robust constraints on the parameter space of particle physics models for dark matter~\cite{Benito:2016kyp}. By employing an angular power spectrum analysis, one can place constraints that are particularly stable with respect to the density profile. This provides a robust assessment of exclusion limits or future sensitivities~\cite{Aartsen:2014hva,Dekker:2019gpe}. It also takes into account the contribution of extra-galactic dark matter annihilation.
We investigate the potential sensitivity of KM3NeT, based on expected experimental performance and detector geometry defined in ref. \cite{Adri_n_Mart_nez_2016} for the the KM3NeT-ARCA configuration, to derive model-independent bounds for various dark matter annihilation channels into SM particles. These sensitivities are the basis to elaborate on the particle physics implications for the dark matter searches and complementary studies.

The second part of this paper focuses on the phenomenological inventory of different models where neutrino telescopes would likely be the most constraining in the future. These models predict naturally prompt neutrinos as SM final states, which are the most sensitive channels for neutrino telescopes. In the simplest scenarios that extend the SM, \ie with only a dark matter particle and a mediating particle, neutrino lines are unavoidably accompanied by charged leptonic final states. The latter produce considerable gamma-ray emission, which is constrained by Fermi-LAT upper limits on the flux of high energy photons from dwarf spheroidal galaxies~\cite{Fermi-LAT:2016uux}.
For an overview of relevant models see \eg~\cite{Lindner_2010,ElAisati:2017ppn}. Here, we consider a subset of the minimal models investigated in ref.~\cite{ElAisati:2017ppn} and perform a global analysis in the light of the future KM3NeT-like neutrino telescopes. Our model building procedure is motivated by the requirements of unitarity, gauge invariance and anomaly free, see \eg the discussion in refs.~\cite{Kahlhoefer:2015bea,Ellis:2017tkh}, and builds on top of the simplest realisation of single mediator models to end up with the known gauged $L_{\mu-\tau}$ model~\cite{He:1990pn,He:1991qd,Foot:1994vd,Baek:2001kca,Ma:2001md,Heeck:2011wj,Altmannshofer:2014cfa}. 
By incorporating different dark matter searches, we examine how the limits for KM3NeT, forecasted with the angular power spectrum analysis, will complement our understanding of dark matter and its nature in terms of annihilation to leptons and ultimately neutrino signals. In this study, we consider Dirac dark matter, which has the potential of the most promising signatures for neutrino telescopes when interacting with SM leptons. For all scenarios, we investigate the relevant parameter space, demonstrating that the future KM3NeT neutrino experiment will play a dominant role in constraining dark matter models, complementary to direct and indirect detection, especially in the case of the $L_{\mu-\tau}$ model.

\section{Angular Power Spectrum Analysis Method}\label{sec:method}

The angular power spectrum (APS) is a powerful probe to asses anisotropies of the neutrino sky and, as demonstrated in previous studies~\cite{Aartsen:2014hva,Dekker:2019gpe}, it can be exploited to firmly test dark matter signals and to place solid constraints on dark matter properties. We here extend the forecast analysis discussed in ref.~\cite{Dekker:2019gpe} to lower dark matter masses (from 200~GeV to $10^5$~GeV) for the KM3NeT neutrino telescope. Being located in the Northern hemisphere, KM3NeT has high sensitivity towards the galactic center with a large field of view, therefore is very suitable to probe for dark matter neutrino signals in our own galaxy and beyond. In~\cref{sec:basics} we describe the neutrino flux expected from annihilating dark matter particles. Then, in~\cref{sec:aps} we detail the angular power spectrum analysis of simulated neutrino skymaps, and in~\cref{sec:bounds} we report the projected model-independent bounds on the dark matter annihilation cross section.
 
\subsection{Dark matter annihilation signals}\label{sec:basics}

Dark matter particles accumulated in the galactic halo can produce a detectable neutrino flux through their annihilation into SM particles. Such a galactic neutrino flux takes the following differential expression
\begin{equation}
    \frac{\mathrm{d} \Phi_{\nu_\beta+\bar{\nu}_\beta}^{\rm gal.}}{\mathrm{d} E_\nu \mathrm{d} \Omega} = \frac12 \frac{\left< \sigma v \right>}{4\pi \, m_{\rm DM}^2} \frac{{\rm d}N_\beta}{{\rm d}E_\nu} \int_0^\infty \mathrm{d}s \, \rho_{\rm DM}^2\left[r\left(s,\ell,b\right)\right]\,,
    \label{eq:gal}
\end{equation}
where $m_{\rm DM}$ and $\left< \sigma v \right>$ are, respectively, the mass and the thermally averaged annihilation cross-section of dark matter, ${\rm d}N_\beta/{\rm d}E_\nu$ is the neutrino energy spectrum per dark matter annihilation, and $\rho_{\rm DM}(r)$ is the dark matter halo density profile as a function of the galactocentric radial coordinate $r=\sqrt{s^2 + R_\odot^2 - 2 s R_\odot \cos\ell\cos b}$ with $R_\odot=8.5~\mathrm{kpc}$ ($b$ and $\ell$ are the galactic angular coordinates). In order to estimate the effect of the choice of the dark matter density profile, we study different distributions. We consider the commonly-used Navarro-Frenk-White (NFW) distribution 
\begin{equation}
    \rho_{\rm DM}^{\rm NFW}(r) = \frac{\rho_0}{\frac{r}{r_s}\left(1 + \frac{r}{r_s}\right)^2}\,,
    \label{eq:nfw}
\end{equation}
with $r_s = 20~\mathrm{kpc}$ and $\rho_0 = 0.33~{\rm GeV/cm^3}$~\cite{Catena:2009mf} (see also ref.~\cite{Pato:2015dua}), and the Burkert profile
\begin{equation}
    \rho_{\rm DM}^{\rm Burkert}(r) = \frac{\rho_0}{\left(1+\frac{r}{r_s}\right)\left[1 + \left(\frac{r}{r_s}\right)^2\right]}\,,
    \label{eq:bur}
\end{equation}
with $r_s = 9.26~\mathrm{kpc}$ and $\rho_0 = 1.72~{\rm GeV/cm^3}$~\cite{Nesti:2013uwa}. While the former provides a large enhancement of the dark matter signal towards the galactic center, the latter predicts a much lower density of dark matter particles in the inner regions of our galaxy. This behaviour typically leads to much weaker dark matter constraints in standard indirect dark matter searches when assuming the Burkert profile (see \eg refs.~\cite{Albert:2016emp,ANTARES:2019svn,Aartsen:2020tdl}). As will be clearly shown later, our forecast analysis based on the angular power spectrum method is instead very feebly affected by the choice of the dark matter galactic distribution.

In order to set model-independent limits on dark matter properties, we examine different phenomenological scenarios where dark matter particles are assumed to have only one annihilation channel at a time with a $100\%$ branching ratio into that final state. For each annihilation channels, the neutrino energy spectrum is computed by using the \texttt{PPPC4DM} package~\cite{Cirelli:2010xx}. These energy spectra take into account the electroweak corrections, whose contribution is relevant for $m_\mathrm{DM} \gg \mathcal{O}(100~\mathrm{GeV})$~\cite{Ciafaloni:2010ti}. It is worth mentioning that an alternative calculation of dark matter spectra with a different treatment of these corrections has been recently provided by the package \texttt{HDMSpectra}~\cite{Bauer:2020jay}. A special case is represented by the annihilation channels into a couple of light bosons which subsequently decay into four neutrinos, $\mathrm{DM}+\mathrm{DM} \rightarrow 4 \nu$. For this channel, since the electroweak corrections have not been included yet, we rely on the analytical box-shaped spectrum for unpolarized light bosons reported in ref.~\cite{ElAisati:2017ppn} (see also refs.~\cite{Ibarra:2012dw,Garcia-Cely:2016pse}). In the next section, we will combine the model-independent upper limits to constrain the parameter space of specific particle physics models, which predict definite branching ratios among the different annihilation channels.

In addition to the galactic component, annihilating dark matter particles would also give rise to a diffuse extragalactic neutrino signal, resulting in neutrino emissions at different redshifts $z$. The extragalactic flux is given by
\begin{equation}
\frac{\mathrm{d} \Phi_{\nu_\beta+\bar{\nu}_\beta}^{\rm ext.gal.}}{\mathrm{d} E_\nu \mathrm{d}\Omega} = \frac12 \frac{\left< \sigma v \right> \, \left(\Omega_{\rm DM}\rho_c\right)^2}{4\pi \, m_{\rm DM}^2} \int_0^\infty \mathrm{d}z \, \frac{B\left(z\right) \, \left(1+z\right)^3}{H\left(z\right)}\left.\frac{{\rm d}N_\beta}{{\rm d}E'_\nu}\right|_{E'_\nu=E_\nu\left(1+z\right)}\,,
\label{eq:extgal}
\end{equation}
where $\rho_c = 5.5\times 10^{-6} \, {\rm GeV \, cm^{-3}}$ is the critical density and $H(z)=H_0 \sqrt{\Omega_\Lambda + \Omega_{\rm m}(1+z)^3}$ is the Hubble expansion parameter with $H_0 = 67.4 \,{\rm km \, s^{-1} \, Mpc^{-1}}$, $\Omega_\Lambda=0.685$, $\Omega_\mathrm{m}=0.315$, $\Omega_\mathrm{DM}=0.264$~\cite{Aghanim:2018eyx}. In the integral over the redshift, the quantity $B\left(z\right)$ is the boost factor (or clumpiness factor) for which we consider the semi-analytical model described in ref.~\cite{Hiroshima:2018kfv} (see ref.~\cite{Ando:2019xlm} for a recent review). The diffuse extragalactic flux is in general neglected in standard indirect dark matter analyses. However, depending on the dark matter halo density profile and on the boost factor, it could further reduce the enhancement of the dark matter signal towards the galactic center and, consequently, weaken the constraints on dark matter annihilation cross-section. In contrast to the studies~\cite{Abbasi:2011eq,Adrian-Martinez:2015wey,Aartsen:2017ulx,Albert:2016emp,Aartsen:2020tdl}, we follow a more conservative approach and include the extragalactic diffuse neutrino flux.

Hence, the total dark matter flux of neutrinos with flavour $\alpha$ reaching the Earth is equal to
\begin{equation}
    \frac{\mathrm{d} \Phi_{\nu_\alpha+\bar{\nu}_\alpha}^{\rm DM}}{\mathrm{d} E_\nu \mathrm{d} \Omega} = \sum_{\alpha \beta} P_{\alpha \beta}\left[ \frac{\mathrm{d} \Phi_{\nu_\beta+\bar{\nu}_\beta}^{\rm gal.}}{\mathrm{d} E_\nu \mathrm{d} \Omega} + \frac{\mathrm{d} \Phi_{\nu_\beta+\bar{\nu}_\beta}^{\rm ext.gal.}}{\mathrm{d} E_\nu \mathrm{d} \Omega} \right]\,,
    \label{eq:DM_flux}
\end{equation}
where $P_{\alpha \beta} = \sum_{i=3}^3 \left|U_{\alpha i}  \right|^2 \left|U_{\beta i}\right|^2$ are the flavour-transition probabilities averaged over cosmological distances. We determine the element of the PMNS mixing matrix $U$ from the recent global neutrino fit discussed in refs.~\cite{Capozzi:2018ubv,Capozzi:2017ipn}.

\subsection{Neutrino skymaps}\label{sec:aps}

We simulate neutrino skymaps with expected dark matter signal and background events for 10 years of data-taking period of the KM3NeT-ARCA site, dedicated to detection of high energy neutrino events as described in ref.~\cite{Adri_n_Mart_nez_2016}, considering only through-going track events in order to reduce the atmospheric muon background. We adopt the flux model from ref.~\cite{Honda:2015fha} describing the background events which are the atmospheric neutrinos that are produced by the interactions of high energy cosmic rays in the atmosphere. The astrophysical flux is subdominant with respect to the atmospheric neutrino flux in the energy range we are interested in (below 100~TeV), hence we neglect any astrophysical contribution to the total skymaps. We refer to null hypothesis for background-only and alternative hypothesis composed of signal plus background events in the following text. 

We bin the neutrino events into \texttt{HEALPix} skymaps using the software package \texttt{HEALPY}~\cite{Gorski_2005}. 
The number of expected neutrino events originating from dark matter signal with a flux $\Phi_{\nu_\mu+\bar{\nu}_\mu}$ (see~\cref{eq:DM_flux}) coming from a region on the sky $\Delta \Omega$ at position $\theta$ (declination) and $\Phi$ (right ascension) is derived by
\begin{equation} \label{eq:Number}
    N_{\nu} \left(\theta, \phi \right)  = p_\mathrm{track} \, \int_{\Delta \Omega} \,d\Omega \, \int_{E_{\rm th}}^{E_{\rm max}} \, dE_{\nu} \, 
    \frac{d\Phi_{\nu_\mu+\bar{\nu}_\mu}}{dE_{\nu}\,d\Omega} \, \mathcal{E}(E_{\nu}, \Omega)\, \rm{vis}(\Omega)\,,
\end{equation}
where $\mathcal{E}(E_{\nu}, \Omega) = T_{\rm obs}\, A_{\rm eff}(E_{\nu}, \Omega)$ is the detector's exposure with $T_{\rm obs}$ the exposure time and $A_{\rm eff}$ the detector's effective area for through-going muon neutrinos. In order to quantify the fraction of the year during which a point on the sky can be observed by KM3NeT, we multiply the function with the visibility function $\rm{vis}(\Omega)$~\cite{Adrian-Martinez:2016fdl}. Only a fraction of muon neutrinos produce track-like events through charged current interactions. To account for this, we further multiply eq.~\eqref{eq:Number} by the probability that charge current interactions take place, which is given by a factor of $p_\mathrm{track} \simeq 0.75$~\cite{Gandhi:1998ri}. In order to gain a larger sensitivity, we integrate only over a small energy range around the peak of the dark matter spectrum with $\rm E_{\nu} = [\frac{1}{10} \rm m_{DM}, \rm m_{DM}]$. The neutrino lines show a sharper peak with respect to other channels, therefore we consider a narrower energy range for this case with $\rm E_{\nu} = [\frac{1}{2} \rm m_{DM}, \rm m_{DM}]$.

\begin{figure}[t!]
\centering
\includegraphics[width=.49\textwidth]{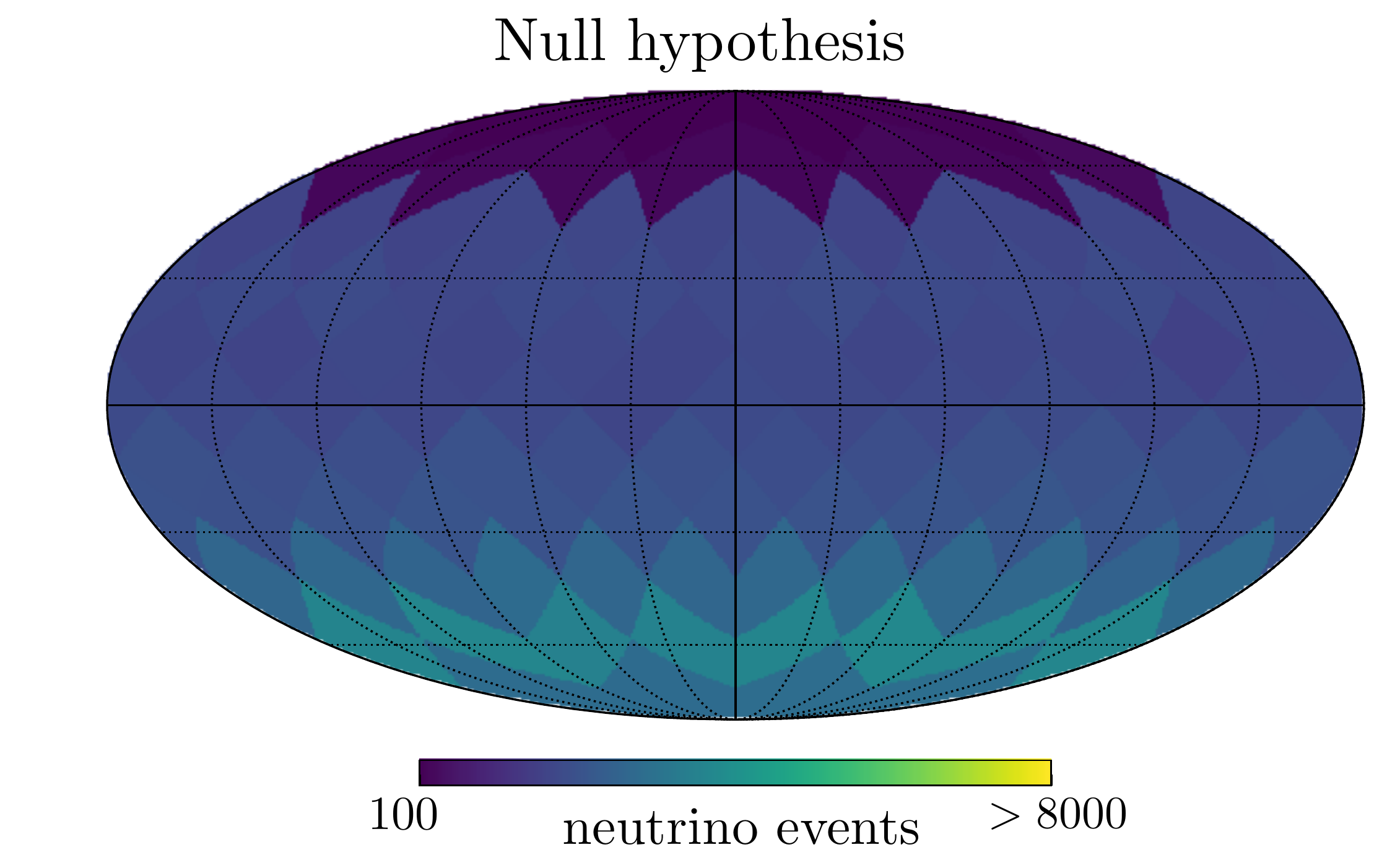}
\includegraphics[width=.49\textwidth]{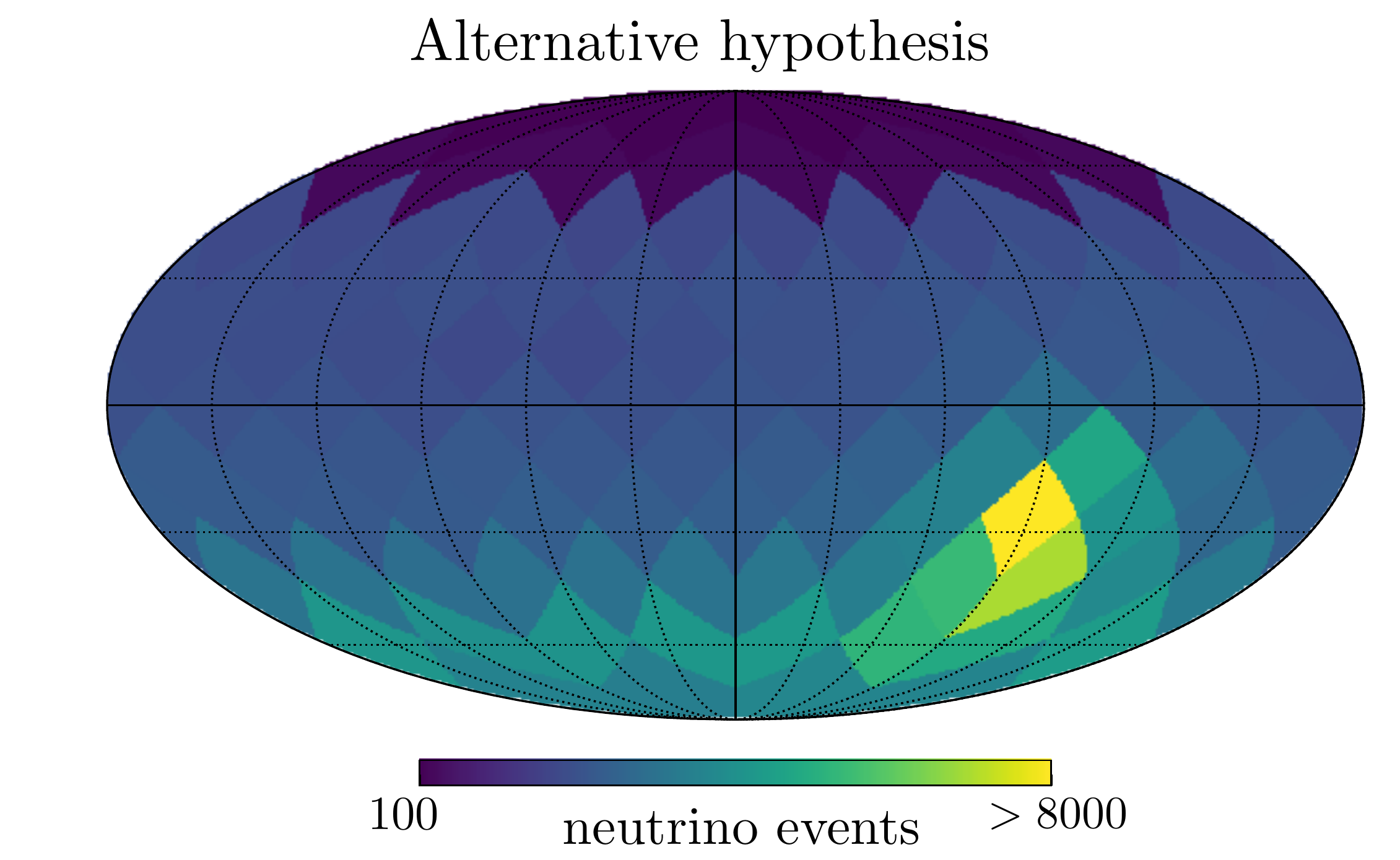}
\caption{\textbf{Left:} Simulated neutrino skymaps for through-going muon neutrinos under the null hypothesis with background-only  for expected 10 years of KM3NeT-ARCA data-taking period. We consider the dark matter annihilation channel of $\mathrm{DM}+\mathrm{DM} \rightarrow \mu^+\mu^-$ with a fixed annihilation cross-section of $\langle \sigma v\rangle = 10^{-22} \, \rm cm^3/s$. \textbf{Right:} Same as left for  the alternative hypothesis including the dark matter neutrino flux. Both skymaps are in equatorial coordinates, and the brightest pixel in the right plot corresponds to the galactic center.}
\label{fig:Skymaps}
\end{figure}

The skymaps consist of isotropic and anisotropic features, where the anisotropic contribution essentially arises from galactic dark matter events. The extent of anisotropy therefore depends on dark matter parameters such as the density profile and annihilation cross-section. We aim at constraining such parameters by analyzing the anisotropic features with respect to the nearly isotropic null hypothesis.
Figure~\ref{fig:Skymaps} illustrates two Monte Carlo simulations for skymaps under the null hypothesis (left) and alternative hypothesis (right) for the $\mu^+\mu^-$ channel with $\langle \sigma v \rangle = 10^{-22} \, \rm cm^3/s$, and for 10 years of expected KM3NeT-ARCA exposure. 
The non-isotropic distribution in the null hypothesis comes from the fact that we consider through-going events with a cut in the zenith angle of the detector, while the skymap is in equatorial coordinates. As it can be seen, we observe a distinguishable anisotropic feature in the right figure originating from the galactic dark matter events. We analyse these patterns through the angular power spectrum, as described below. 

The APS describes the fluctuations as a function of angular scale. The skymap is expanded into spherical harmonics through
\begin{equation}
N_{\nu}\left(\theta, \phi \right) = \sum_{\ell m} a_{\ell m} Y_{\ell m}\left(\theta, \phi \right) \,,
\end{equation}
where $Y_{\ell m}\left(\theta, \phi \right)$ are the spherical harmonic functions and $a_{\ell m}$ the expansion coefficients. The APS is described by the average of the expansion coefficients over the sky
\begin{equation}
C'_{\ell} = \frac{1}{2\ell +1} \sum_{m=-\ell}^{\ell} |a_{\ell m}|^2 \,.
\end{equation}
To compute the APS, we use the numerical function \texttt{anafast} from the software package \texttt{HEALPix}~\cite{Gorski_2005}. We are interested in anisotropic features only, and in order to discard any information on the total number of events, we remove the monopole and normalize the coefficients as $C_{\ell}=C'_{\ell}/\rm N_{tot}^2$. The largest contributions to the APS come from the first multipole moments. Therefore, we analyse the APS with maximum moment $\ell_{\rm max}=8$. We do this despite the fact that KM3NeT's angular resolution allows one to go higher.  
For each dark matter model, we perform $10^5$ Monte Carlo simulations, vary the cross-sections between $\left<\sigma v\right>=[10^{-25},10^{-22}]\,\rm{cm^3 s^{-1}}$ in steps of $\Delta\log_{10}\langle\sigma v\rangle = 0.2$, and calculate the corresponding APS. 
Additionally, we generate mock data sets by performing $10^5$ Monte Carlo simulations under the background-only hypothesis. In order to have a statistical measure for the goodness of the models being tested, we apply the following $\chi^{2}$,
\begin{equation}
\chi^{2} \left(C_{\ell}\right) = \sum_{\ell \ell'} \left(C_{\ell} - C_{\ell}^{\rm{mean}}\right) (\rm{Cov}_{\ell\ell'})^{-1} \left(C_{\ell'} - C_{\ell'}^{\rm{mean}}\right) \,,
\label{eq:chi2}
\end{equation}
where $C_{\ell}$ is the APS of one simulation, $C_{\ell}^{\rm{mean}}$ is the mean value and $\rm{Cov}_{\ell\ell'}$ is the covariance matrix, where $C_{\ell}^{\rm{mean}}$ and $\rm{Cov}_{\ell\ell'}$ are obtained from a complete set of simulation under the alternative hypothesis. For each characterization of the model, we calculate the probability density function of $\chi^2$, $P(\chi^2|\Theta)$ with $\Theta$ the being the set of the parameters describing the dark matter signal component. We compute $\chi_{\rm md}^2\equiv \chi^2(C_{\ell}^{\rm{md}})$ from the mock data (md) sets under the background-only hypothesis in order to obtain the probability of having the same or more extreme values of $\chi^{2}$ by the following $p$--value,
\begin{equation}
p = \int_{\chi^2_{\rm md}}^\infty d\chi^2 P(\chi^2|\Theta).
\end{equation}

We simulate $10^5$ background-only skymaps corresponding to the Asimov data sets in order to obtain the  $p$--value distribution \cite{Cowan_2011}. From this distribution we derive the median, as well as the $2\sigma$ contour band. The expected upper-limit on dark matter annihilation cross-section for a given channel at a certain mass value is then obtained at 90\% confidence level (CL) at $p \leq 0.10$.

\subsection{Model-independent bounds}\label{sec:bounds}

\begin{figure}[t!]
    \centering
    \includegraphics[width=\textwidth]{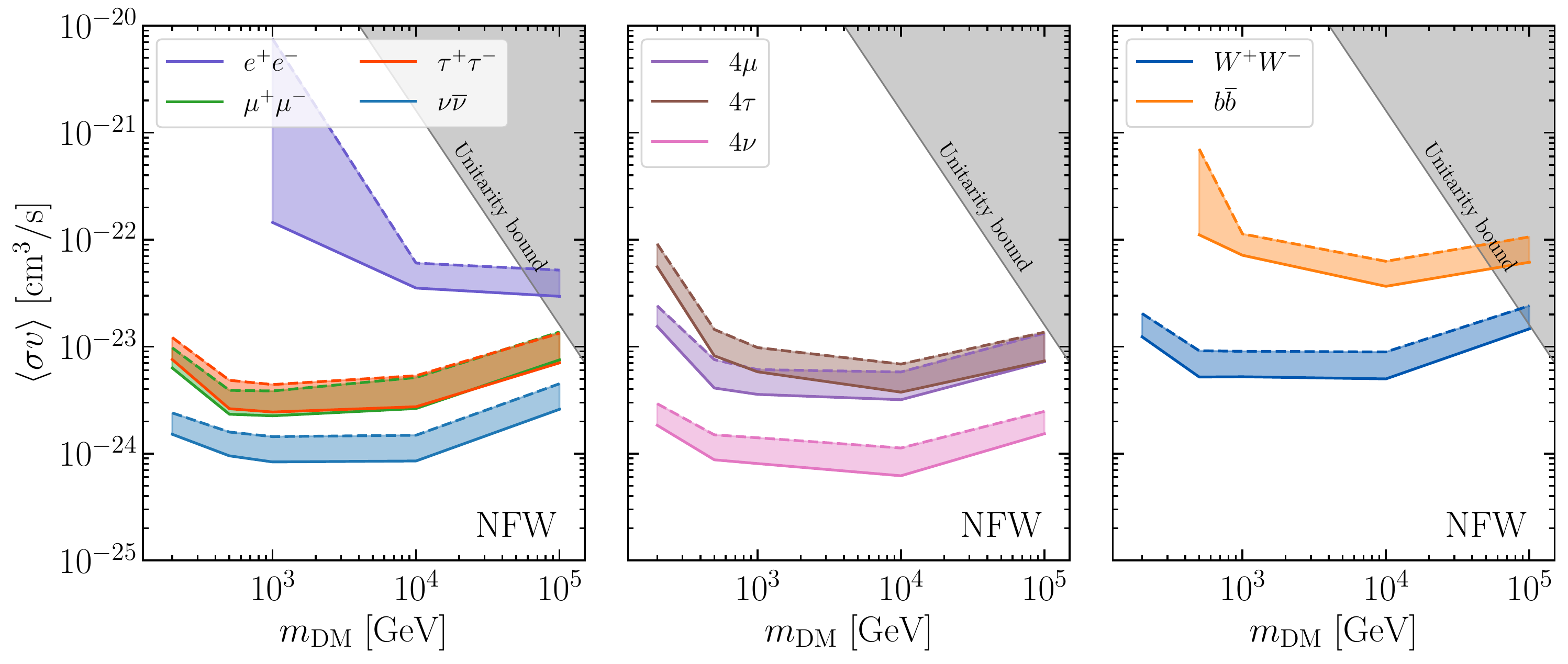}
    \caption{\textbf{Left:} Forecasted upper-limits at 90\% CL to the dark matter annihilation  cross-section $\langle \sigma v \rangle$ as a function of dark matter mass $m_\mathrm{DM}$, for 10-year exposure of KM3NeT-ARCA using the NFW halo density profile. Annihilation with branching ratio 100\% into a pair of leptons is considered, as labelled by the different colours. The bands represent the median (solid lines) and conservative $2\sigma$ (dashed lines) upper-limits obtained from the Monte Carlo simulations. The grey region is excluded by unitarity. \textbf{Centre and right:} Same as left for the 4 leptons final state and for representative SM final states into quarks and gauge bosons respectively.
        }
    \label{fig:constraints}
\end{figure}

Here, we present the results obtained through the angular power spectrum forecast analysis as described above in detail. In~\cref{fig:constraints}, we report the future KM3NeT sensitivity at 90\% CL with 10-year exposure to WIMP dark matter annihilation cross-section for different channels. For the channels involving neutrinos in the final states, we assume equipartition among neutrino flavours. Nevertheless, neutrino channels with a specific flavour at the production result in very similar bounds due to neutrino oscillations during the propagation to the Earth. In the plots, the bands have been obtained by analyzing the angular power spectrum of $10^5$ simulated neutrino skymaps for each value of the dark matter mass considered. In particular, the solid and dashed lines represent the median and conservative $2\sigma$ upper bounds obtained from the Monte Carlo simulations at 90\% CL, assuming the NFW halo profile for the galactic dark matter distribution. The grey region is excluded by the requirement of unitarity of the dark matter annihilation cross-section~\cite{Griest:1989wd}. As can be seen in the plots, the limits for the electron and bottom quark channels stop at $m_\mathrm{DM} = 1~\mathrm{TeV}$ and $m_\mathrm{DM} = 500~\mathrm{GeV}$, respectively. Below such masses, the detection efficiency is very suppressed for both channels. In case of the electron channel, neutrinos are only produced through the electroweak radiation which are not very efficient below TeV energies producing detectable neutrinos in our study. For dark matter masses below 600 GeV, most of the neutrinos, produced by the annihilation into bottom quarks, have an energy smaller than $100~\mathrm{GeV}$, which is below the sensitivity of the KM3NeT-ARCA detector. This is why the upper limits break down at that dark matter mass.

\begin{figure}[t!]
    \centering
    \includegraphics[width=\textwidth]{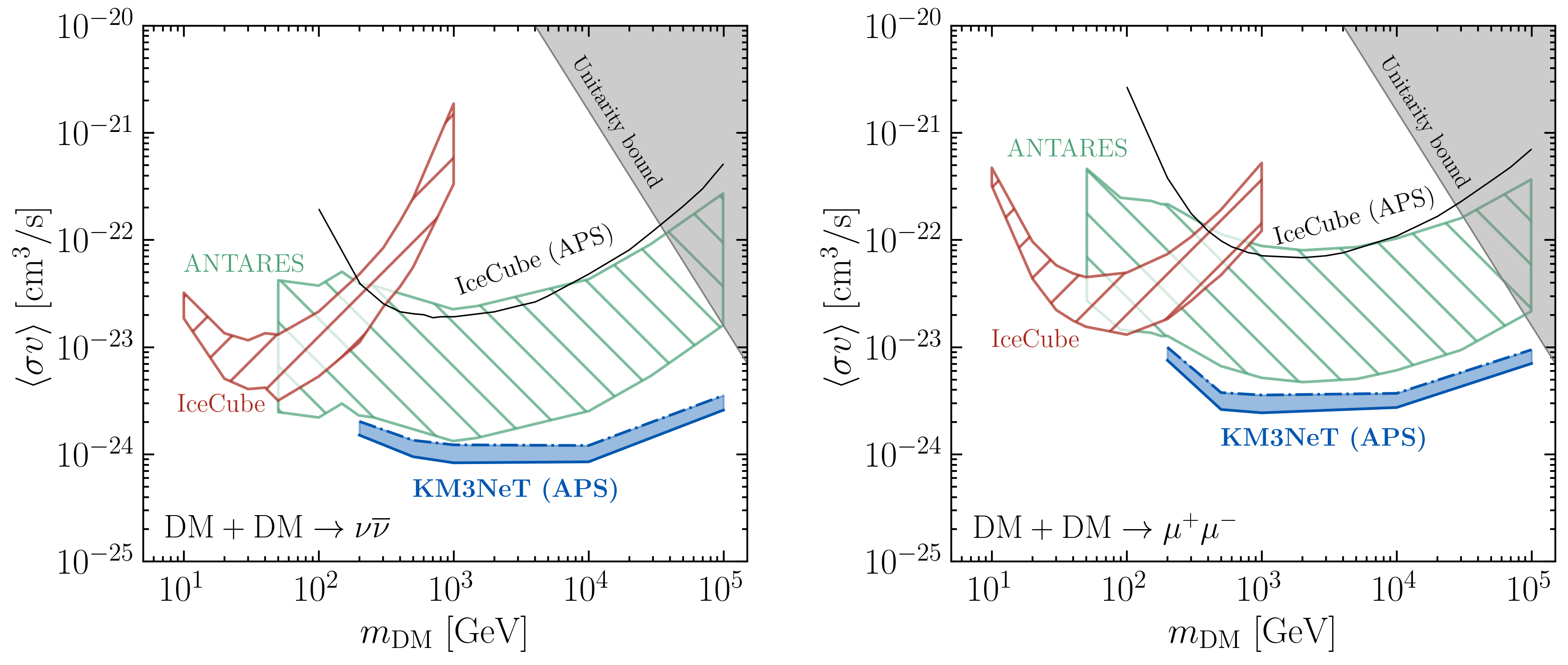}
    \caption{\textbf{Left:} Present and future limits placed by neutrino telescopes on the dark matter annihilation cross-section to neutrinos. The bands represent the uncertainty related to different galactic dark matter profile: the lower (upper) edges refer to NFW (Burkert) density profile. The blue region is the median sensitivity at 90\% CL after 10-year exposure of KM3NeT obtained through the angular power spectrum (APS) method in case of NFW (solid lines) and Burkert (dot-dashed lines) profiles. The other bands corresponds to the dark matter search analyses: 3-year IceCube~\cite{Aartsen:2017ulx} (red band), 11-year ANTARES~\cite{ANTARES:2019svn} (green band), and 1-year IceCube with a similar multipole study~\cite{Aartsen:2014hva} (black band). The grey region in the top-right corner is excluded by unitarity (see text). \textbf{Right:} Same as left for dark matter annihilation into muons.}
    \label{fig:density}
\end{figure}

In~\cref{fig:density} we graphically quantify the systematic uncertainty affecting our constraints (blue band) according to different choices for the dark matter distribution in our galaxy. In particular, we show the median sensitivity at 90\% CL for the two extreme cases of NFW (solid blue lines) and Burkert (dot-dashed blue lines) profiles for two annihilation channels. The plots correspond to two different annihilation channels of dark matter particles: neutrino lines (left plot) and charged muons (right plot). We also report the existing upper limits placed by dark matter searches in neutrino telescopes: 3-year IceCube~\cite{Aartsen:2017ulx} (red band), 11-year ANTARES~\cite{ANTARES:2019svn} (green band), and 1-year IceCube with a similar multipole analysis study~\cite{Aartsen:2014hva} (black band) simply denoted as ``IceCube APS'' for the sake of brevity. For consistent comparison, we have properly scaled these constraints to our set of dark matter halo parameters (see eqs.~\eqref{eq:nfw} and~\eqref{eq:bur}).
 
One can observe that the APS method is very stable over different halo profiles. On average the Burkert distribution weakens the limits by only $\sim 40\%$ with respect to NFW. For example, taking $m_\mathrm{DM} = 10^4~\mathrm{GeV}$, annihilation cross-section into neutrinos will be constrained to be below $1.21 \times 10^{-24}~{\rm cm^3/s}$ and $8.52 \times 10^{-25}~{\rm cm^3/s}$ for Burkert and NFW respectively. Comparing this result to studies which do not use the APS method such as ANTARES~\cite{ANTARES:2019svn}, there is a much greater variation due to the choice of halo profile. Moreover, if we compare our results to that of another APS study, such as the one performed for IceCube in ref.~\cite{Aartsen:2014hva}, we see a projected improvement of over an order of magnitude. Obtaining these robust limits will have significant implications on the particle physics interpretations. In the next section we elaborate on this.

\section{Dark Matter Models for Neutrino Telescopes}\label{sec:models}

While in the previous section we have presented model-independent bounds for the future KM3NeT-like telescope, in this section we provide interpretation of these bounds in terms of selected minimal dark matter models, which have the advantage that KM3NeT will give the most competitive insight for. Additionally, interpretation of bounds allows for a complementary analysis that compares the sensitivity of different dark matter probes. In~\cref{subsec:constraints} we outline the model characteristics and the other experimental constraints we consider in this study. Subsequently, we investigate the parameter space with different variables of models, which will be explained in~\cref{subsec:tchan}, in~\cref{subsec:schan} and in~\cref{subsec:gaugedzp}.

\subsection{Minimal models for neutrino signals}

From the model building perspective, our starting point is the simplified model framework, which has been investigated recently at the LHC~\cite{Abercrombie:2015wmb,Boveia:2016mrp,Abdallah:2015ter,Kahlhoefer:2017dnp,Arcadi:2017kky,Arina:2018zcq}. This entails making minimal additions to the SM by way of couplings and particles in order to incorporate a dark matter particle candidate. It allows for the exploration of scenarios in a rather model-independent way and eases the comparison of theoretical predictions with the various experiments. Such models can be categorised in terms of $s$-channel and $t$-channel, which from a structural point of view are very different. 

The $s$-channel models feature a new boson which mediates between two dark matter and two SM particles, hence it is even under the dark group that protects the dark matter particle from decaying. The new boson is also a singlet under the SM gauge group. As a consequence, this mediator can be lighter or heavier than the dark matter.

Alternatively, $t$-channel models feature a new interaction vertex coupling directly one dark matter particle with a new  mediator and a SM particle. This requires a discrete $Z_2$ and continuous global $U(1)$ symmetries to stabilise the real and complex dark matter candidate respectively. It also implies that the mediator is charged under the dark group, hence can only be heavier than the dark matter particle.

More specifically, among the various realizations of $s$- and $t$-channel models, we select models that share at least these two properties:
\begin{description}
\item[$\bullet$] \textbf{Leptophilic models:}  The mediator and the dark matter do not couple at tree level to quarks (nor to the Higgs and to electroweak gauge bosons) by construction, but feature purely leptonic final states, preferably neutrinos, which are the most promising for neutrino telescopes when compared with other dark matter searches. In particular, annihilation of dark matter can produce sizable neutrino lines or neutrino box signals at tree level, with branching ratio close or equal to the one into charged leptons, when latter annihilation channels can not be suppressed. 
\item[$\bullet$] \textbf{$s$-wave annihilation}: We require that the dark matter velocity/thermally averaged annihilation cross-section $\sigmav$ into SM particles is independent of the relative dark matter velocity. This means that at present time $\sigmav$ can be potentially large in galactic halos (characterized by $ v ~\simeq \left(10^{-5} - 10^{-3}\right) c$) and in the reach of KM3NeT. 
We do not focus however on thermal dark matter scenarios but consider the whole parameter space, remaining agnostic on the mechanism that provides 100\% of the relic density $\Omega h^2$ measured by Planck~\cite{Aghanim:2018eyx}. Notice that the values to which KM3NeT is sensitive, as obtained in previous section with APS ($\sigmav \simeq 10^{-24} \rm cm^3/s$), denote the region of under-abundant dark matter assuming standard freeze-out, because $\Omega h^2 \propto \sigmav^{-1}$. However we will show that for certain models, KM3NeT will be able to probe thermal values obtained with the standard freeze-out mechanism, see~\cref{subsec:gaugedzp}. 
\end{description}
\begin{figure}
  \centering
  \includegraphics[trim={0 0 0 0} ,clip,width=.32\columnwidth]{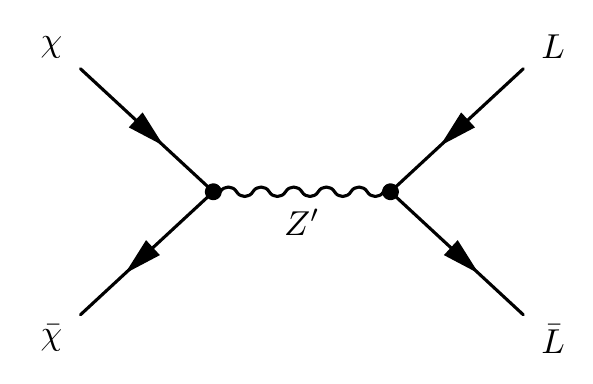}
  \includegraphics[trim={0 0 0 0},clip,width=.32\columnwidth]{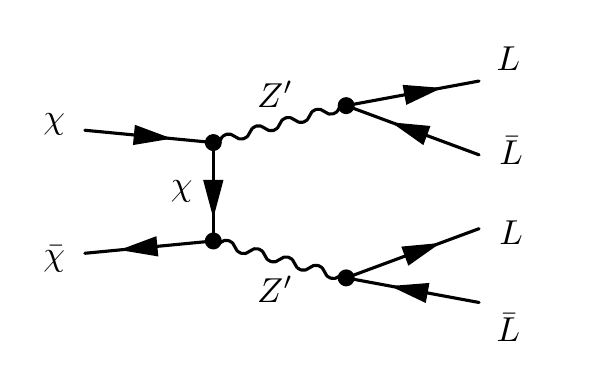}
  \includegraphics[trim={0 0 0 0},clip,width=.32\columnwidth]{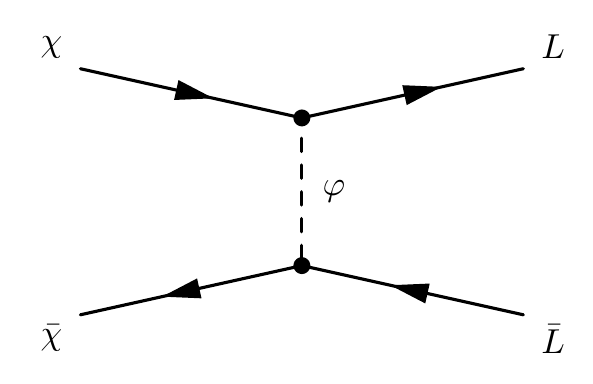} 
  \caption{
    \label{fig:diagrams}
    \textbf{Left and centre:} Representative Feynman diagrams describing the
    annihilation of dark matter particles into leptonic $SU(2)_L$ doublets for the $s$-channel models under study. Notice that in the central diagram the $Z'$ are produced on shell and decay subsequently into leptons. \textbf{Right:} Same as left for the $t$-channel model we consider in our analysis.}
\end{figure}

The latter requirement severely restricts the possibilities in term of $s$- and $t$-channel models. Taking the dark matter to be a singlet Dirac Fermion ($\chi$) the only viable choice for the mediator in case of $s$-channel models is spin 1 ($Z'$ henceforth), while for  $t$-channel is a scalar mediator ($\varphi$). The relevant annihilation diagrams are shown in~\cref{fig:diagrams}. These models are a subset of those studied in~\cite{Lindner_2010,ElAisati:2017ppn} in the context of neutrino line signals. 
 $m_{\chi}\sim 550$ GeV, Fermi limits from the charged lepton channel (blue line) dominate.  

Reference~\cite{ElAisati:2017ppn} exhibited well the difficulty a model builder has when trying to optimise the prospects of neutrino  telescopes. 
This is especially true when one is attempting to work with a simple extension of the SM that is theoretically well-motivated. In the next sections we will build on ref.~\cite{ElAisati:2017ppn} by considering specific coupling configurations and adding vital direct detection constraints. Reference~\cite{Blennow:2019fhy} also study thoroughly neutrino line signals, however
the phenomenology is related to a neutrino portal for dark matter, which is achieved with a right-handed neutrino in addition to dark matter. The mixing between active-sterile neutrinos is what drives the dark matter to SM interaction rate. Here we free ourselves from such considerations and only consider a new dark matter candidate. This allows for interesting regions of parameter space at higher dark matter masses. 

The specifications of each model is described in the following sections as they are introduced, first we discuss all the experimental constraints (indirect detection, direct detection, collider searches and cosmological bounds) we consider in conjunction with the potential of the analysis technique which is described in previous section.

\subsection{Complementary constraints to neutrino telescopes}\label{subsec:constraints}

The models we analyse are implemented in FeynRules~\cite{Alloul:2013bka} and we use the corresponding UFO files to compute $\sigmav$ with \textsc{MadDM}~\cite{Ambrogi:2018jqj} for both the thermally averaged cross-section for relic density and the annihilation cross-section at present time, which actually coincide at leading order as we consider $s$-wave annihilation. Notice that the predicted flux of SM particles, see~\cref{eq:DM_flux}, is further divided by a factor of 1/2 to account for the Dirac nature of the dark matter.

As far as the complementary is concerned, let us start with indirect detection.
As already anticipated above, leptophilic dark matter features as final states charged leptons, which produce gamma rays. One of the most constraining and robust limits for a continuum gamma-ray spectrum is provided by the Fermi-LAT bounds from dwarf spheroidal galaxies (dSphs)~\cite{Fermi-LAT:2016uux}. In order to calculate the limits at 90\% CL for the dSph Fermi-LAT constraint we use the likelihood method implemented within \textsc{MadDM} for determining the exclusion limits given the specific model realisation. 
We adopt the J-factors for ultrafaint dSphs (which were also used in ref.~\cite{Ambrogi:2018jqj}) from ref.~\cite{Ando:2020yyk}, where more realistic assumptions for satellite formation are made to compute the J-factors from stellar kinematic for ultrafaint dSphs.
This has the impact of weakening the Fermi-LAT bounds because ultrafaint dSphs contribute significantly to the exclusion limits. This updated J-factors modify the exclusion limits by roughly a factor of $\sim 4$ for $m_{\chi}\gtrsim 100$ GeV.  

\begin{figure}
    \centering
    \includegraphics[width=0.7\textwidth]{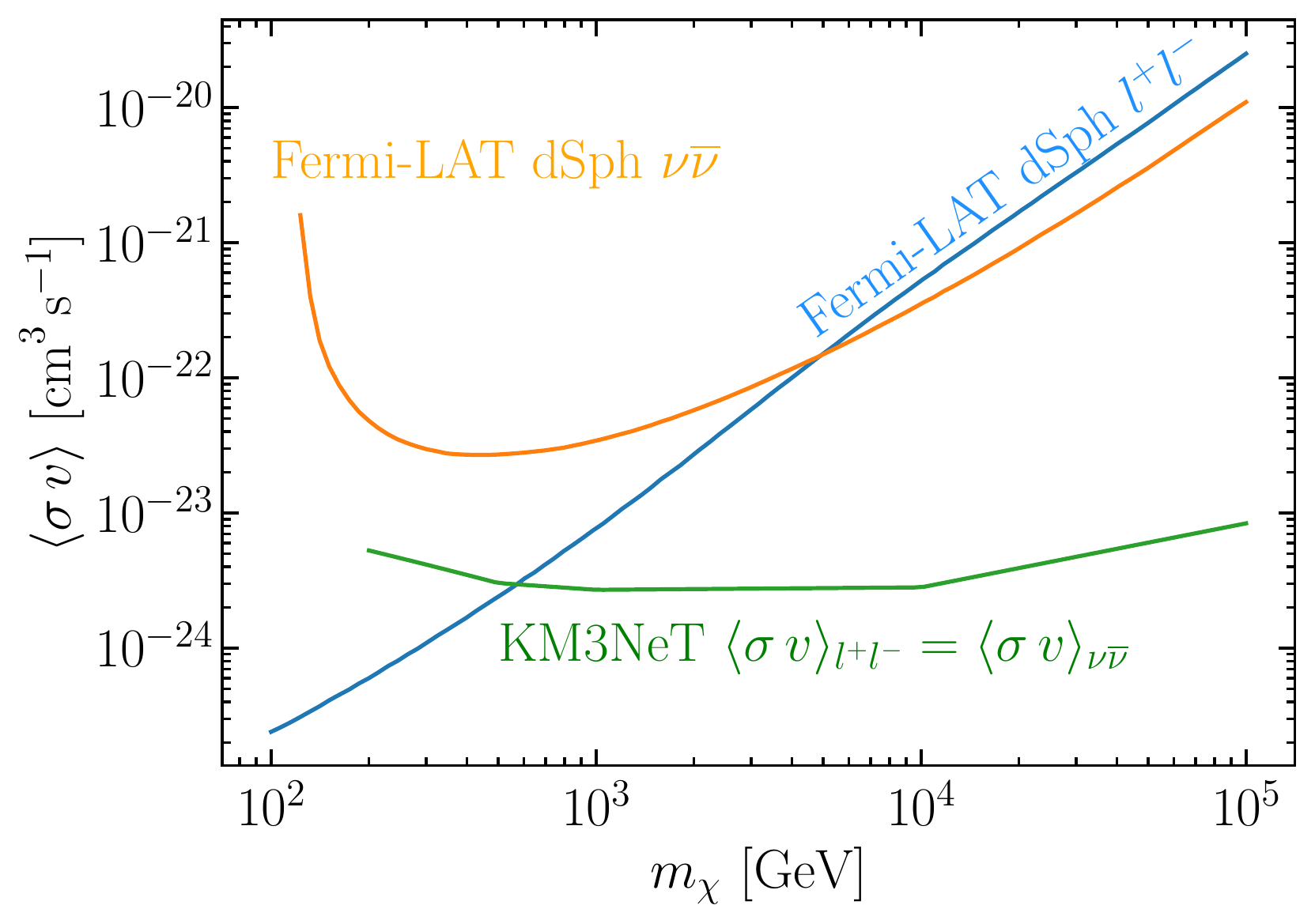}
    \caption{Fermi-LAT exclusion limits recasted with \textsc{MadDM} from dwarf spheroidal galaxies (dSphs)~\cite{Fermi-LAT:2016uux}, with the assumption of $\langle\sigma v\rangle_{l^{\pm}}=\langle\sigma v\rangle_{\nu_{l}}$, in the $\{\sigmav, m_\chi\}$-plane. Charged leptons and neutrino line final states are depicted by blue and orange solid lines respectively. The projected upper-limit for KM3NeT neutrino telescope with the same assumption is shown by the green solid line.}
    \label{fig:Fermi_limit}
\end{figure}

The exclusion limits for the case of annihilation into two body SM particles are based on energy spectra from PPPC4DM~\cite{Cirelli:2010xx} including electroweak corrections implemented as in~\cite{Ciafaloni:2010ti}. These are the same that are used for the KM3NeT analysis, see~\cref{sec:basics}. For  deriving the neutrino box spectra, we use the analytical formula provided in~\cite{ElAisati:2017ppn} and the spectra with 4 leptons in the final state from PPPC4DM, in the case of Fermi-LAT bounds we generate the energy spectra with \textsc{Pythia 8}~\cite{Sjostrand:2014zea}, as implemented within \textsc{MadDM}. However we do not have the handling on the electroweak corrections for the 4 neutrino final state because those are not included in the \textsc{Pythia 8} version released with \textsc{MadDM}.
For dark matter masses well above the TeV scale, electroweak corrections are relevant and change the energy spectra especially of charged leptons and neutrinos. More importantly neutrinos emit electroweak radiation producing hence a secondary flux of gamma rays which can be constrained by gamma ray observations. Our bounds include the contribution of the gamma-ray flux from neutrinos in the Fermi-LAT dSph exclusion limits, similarly to ref.~\cite{Queiroz:2016zwd}, in the case of two neutrino final state. Figure~\ref{fig:Fermi_limit} illustrates that the recasted exclusion limits from gamma rays induced by neutrino final states are subdominant with respect to the gamma-ray exclusion limits coming from charged leptons, when those are present, for most of the relevant dark matter mass range. Intriguingly they overtake charged lepton exclusion limits above 5 TeV masses. This originates from the interplay of two factors: 
\begin{itemize}
\item The energy spectrum of gamma rays induced by electroweak corrected neutrino final states populates with more events the small gamma-ray energy ($E_\gamma$) range; this increases the sensitivity of Fermi-LAT measurements when the dark matter is heavy.
\item The gamma-ray energy spectra from charged leptons take into account only the prompt gamma-ray contribution, which peaks towards large $E_\gamma/m_{\rm DM}$. Since the sensitive energy window of Fermi-LAT is between 300 MeV up to 300 GeV, only the lower part of the energy spectra contribute in the case of heavy dark matter. This low energy part may receive additional contribution by the inclusion of additional secondary gamma rays, produced for instance by Inverse Compton scattering (ICS) processes, see \eg this review~\cite{Blumenthal:1970gc}. It is however not clear how ICS would affect Fermi-LAT bounds for heavy dark matter, as there have not been thorough studies of ICS in dSphs, while radio emission and X-ray emission have been more deeply investigated, see \eg~\cite{Jeltema:2008ax,Natarajan:2015hma,Kar:2019cqo}. Furthermore, the secondary gamma-ray flux relies on additional astrophysical assumptions for the modelling of the interstellar medium, hence the derived upper limits are subject to large uncertainties, similarly to the case of our galactic halo~\cite{Buch:2015iya}.
\end{itemize}

Regardless of these interesting details, we see that above dark matter masses of $\sim 500$ GeV, forecasted limits for KM3NeT show substantial gains over the Fermi-LAT limits.

The models we consider are leptophilic by constructions and naively, one would assume that they avoid all direct dark detection bounds since there is no tree level vertex which couples the quarks or gluons directly to the dark matter particle. However, renormalization effects (RGEs) generate effective couplings to the nuclei which lead to constraints from hadronic processes as shown in refs.~\cite{DEramo:2016gos,DEramo:2017zqw}. To compute the effect of RGEs for the $Z'$ model we use {\sc RunDM}\footnote{\textsc{RunDM} is available here \url{https://github.com/bradkav/runDM}~\cite{DEramo:2016gos}.}, while for the $t$-channel case we evaluate directly the loop diagrams~\cite{Cerdeno:2019vpd}. The details on the specific contributions to the elastic scattering cross section is provided in the pertinent model sections.
All models generate spin-independent couplings, hence we consider the effect of RGEs relating to the XENON1T experimental results~\cite{Aprile:2018dbl}, recasted using the {\sc RAPIDD} tool~\cite{Cerdeno:2018bty}. We additionally discuss the dependence of the XENON1T bound on astrophysical parameters such as the local dark matter density. We consider as reference value $\rho_{\odot} = 0.4 ~\rm GeV/cm^3$~\cite{Bozorgnia:2016ogo}, but exemplify ones the impact of this choice by considering the whole range of allowed values, $\rho_{\odot} = (0.2$--$0.4)~ \rm GeV/cm^3$; see ref.~\cite{Ibarra:2018yxq} and references therein.

There are several constraints from collider experiments that may be relevant for our study. First it should be noted that we are interested in the electroweak production of the new mediator and/or dark matter particles, as the couplings to quarks are negligible by construction. The $t$-channel model can be searched for with pair-production of the charged $\varphi$, which consequently decays into two charged leptons and the dark matter ($pp \to \varphi^+ \varphi^- \to \chi \bar{\chi} l^+ l^-$, with $l=e,\mu$). This resembles to the supersymmetric search for slepton production in the simplified model framework~\cite{Aad:2019qnd}. 
Nevertheless, this search can constrain the mediator masses up to roughly 300 GeV, which is below the mass limits we consider in this study. As far as it concerns $s$-channel models with vector mediator, the most sensitive searches such as mono-X searches, di-leptons, di-quarks do not apply here, because they all assume a non-zero coupling with quarks, see for instance~\cite{Aad:2019fac,Aad:2019hjw,Aaboud:2018fzt,ATLAS:2020jhr,ATLAS:2020wzf}. The strongest constraint for vector mediator mass comes from LEP-II bounds for the process $e^+ e^- \to f \bar{f}$ (where $f$ are the SM fermions)~\cite{Zyla:2020zbs} and states that $m_{Z'} \gtrsim 209 \, \rm GeV$. We conclude that collider bounds are in general fairly limited for leptophilic dark matter models in the ball-park of detection of neutrino telescopes. 

Very light vector mediators are allowed under the assumptions that the coupling to leptons is tiny. Several constraints arise for mediators in the mass range in between MeV and GeV, most notably the search $Z \to 4 \mu$ from BaBar~\cite{TheBABAR:2016rlg} and CMS~\cite{Sirunyan:2018nnz}. All the other constraints arise from neutrino physics, notably neutrino trident production~\cite{Altmannshofer:2014pba}, neutrino-electron scattering in Borexino~\cite{Kaneta:2016uyt} and neutrino cooling of white dwarfs~\cite{Bauer:2018onh}. In our analysis we will consider mediator masses down to 1 MeV, below the muon production threshold due to the two-body decay of the $Z'$. However our $Z'$ will couple very feebly with the SM particles by assumption, with couplings $\mathcal{O}(10^{-10}-10^{-7})$: this  scenario is known as secluded dark matter~\cite{Pospelov:2007mp}. Bounds from collider, neutrino and fixed target experiments can then be easily avoided in such configuration, however there are cosmological bounds which are of interest in the case of very light mediators and will be discussed in section~\ref{subsec:gaugedzp}.

\subsection{Scalar mediated, t-channel annihilation}\label{subsec:tchan}

The simplest realisation of $t$-channel model consistent with $s$-wave annihilation features a singlet Dirac dark matter candidate and a scalar mediator. It is similar to dark matter $t$-channel models currently investigated at the LHC~\cite{Arina:2020udz,Arina:2020tuw}, with the difference that here the mediator is not a coloured particles as it couples to the SM leptons, see \eg~\cite{Ibarra:2015fqa}.\footnote{The model is available in the {\sc FeynRules} database \url{https://feynrules.irmp.ucl.ac.be/wiki/DMsimpt}.} The interaction Lagrangian coupling dark matter to leptons is 
\begin{equation}
\mathcal{L}^{\varphi} = y_{\alpha} \bar{\chi} L_{\alpha} \varphi^{\dagger} + \rm h.c.\,, 
\label{eq:scalar_lag}
\end{equation}
where $\varphi$ is a $SU(2)$ doublet that interacts solely with the left-handed leptonic $SU(2)$ doublet $L_\alpha$, with the $\alpha$ index running over the three leptonic flavour $e, \mu, \tau$. 
The $y_\alpha$ coupling strength is a $3\times 3$ matrix in flavour space, that we take real for simplicity.
As with the Yukawa couplings in the SM, we assume a hierarchical structure among coupling strengths across generations, inspired by the criterion of minimal flavour violation~\cite{DAmbrosio:2002vsn}. The dominant contribution will come from the $\tau$ flavour, $y_\tau$, while we consider $y_e$ and $y_\mu$ smaller and negligible. The coupling is the same for the upper and lower component of the $SU(2)$ left-handed leptonic doublet, hence the branching ratio is the same for annihilation into a pair of neutrinos $\nu_\tau$ or into the pair of $\tau^+\tau^-$. In total, this model has three free parameters:
\begin{equation}
    \{m_{\chi}, m_{\varphi}, y_{\tau} \}\,.
\end{equation}
Concerning the dark matter mass range, we explore the most favourable mass range for KM3NeT-ARCA, as established in the previous section. 
The mediator mass is scanned over in the same mass range with the constraint of being always heavier than the dark matter mass, at least by a tiny amount, $(m_\varphi/m_\chi - 1) \gtrsim 0.1$. We perform a scan for two values of the coupling strength, $y_\tau =\sqrt{4 \pi}$, and the maximal value allowed by perturbativity, which is $4\pi$. 

Reference~\cite{ElAisati:2017ppn} explores this interaction, as well as slightly more complicated realizations. We extend their work on this simplest case by considering the constraints coming from direct detection. At loop level, this model provokes interactions between dark matter and photons~\cite{Kopp:2014tsa,Ibarra:2015fqa}. At the energy scale of direct detection, these interactions are parameterised and constrained by non-renormalisable effective operator vertices~\cite{Ibarra:2015fqa,Kavanagh:2018xeh,Arina:2020mxo}, 
\begin{eqnarray}
\mathcal{L}_{\chi\chi\gamma}  &\supset& \frac{\mathcal{C}_{\mathcal{M}}}{2 \Lambda} \bar{\psi} \sigma^{\mu \nu} \psi \cdot B_{\mu \nu} +\frac{\mathcal{C}_{el}}{2 \Lambda} i \bar{\psi} \sigma^{\mu \nu} \gamma^{5} \psi \cdot B_{\mu \nu} \nonumber \\
&\hspace{1em}& +\frac{\mathcal{C}_{cr}}{\Lambda^{2}} \bar{\psi} \gamma^{\mu} \psi \cdot \partial^{\nu} B_{\mu \nu} + \frac{\mathcal{C}_{\mathcal{A}}}{\Lambda^{2}} \bar{\chi} \gamma^{\mu} \gamma^{5} \chi \cdot \partial^{\nu} B_{\mu \nu}
\label{eq:moments_lag}.
\end{eqnarray}
Here, $B_{\mu\nu}$ is the hypercharge field strength tensor, $\mathcal{C}_j$ are the dimensionless Wilson coefficients for the dimension-5 operators, which are the electric and magnetic dipole moments $(\mathcal{C}_{el})$ and $(\mathcal{C}_{\mathcal{M}})$, and for the dimension-6 operators, that are the charge radius $(\mathcal{C}_{cr})$ and anapole $(\mathcal{C}_{\mathcal{A}})$ interactions. Of which, the electric dipole interaction is particularly well constrained by recent results from XENON1T~\cite{Aprile:2018dbl}, where limits on $\frac{\mathcal{C}_{el}}{\Lambda}$ are $\sim 10^{-8}$ for $m_{\chi}\sim 1$ TeV. However, for the model at hand, an electric dipole response is only provoked by complex Yukawa couplings, which we are not considering. The next most constraining limits are for the magnetic and charge-radius interactions, which are those included in the XENON1T bounds we compute, as in~\cite{Arina:2020mxo}.

\begin{figure}
	\centering
	\includegraphics[width=\textwidth]{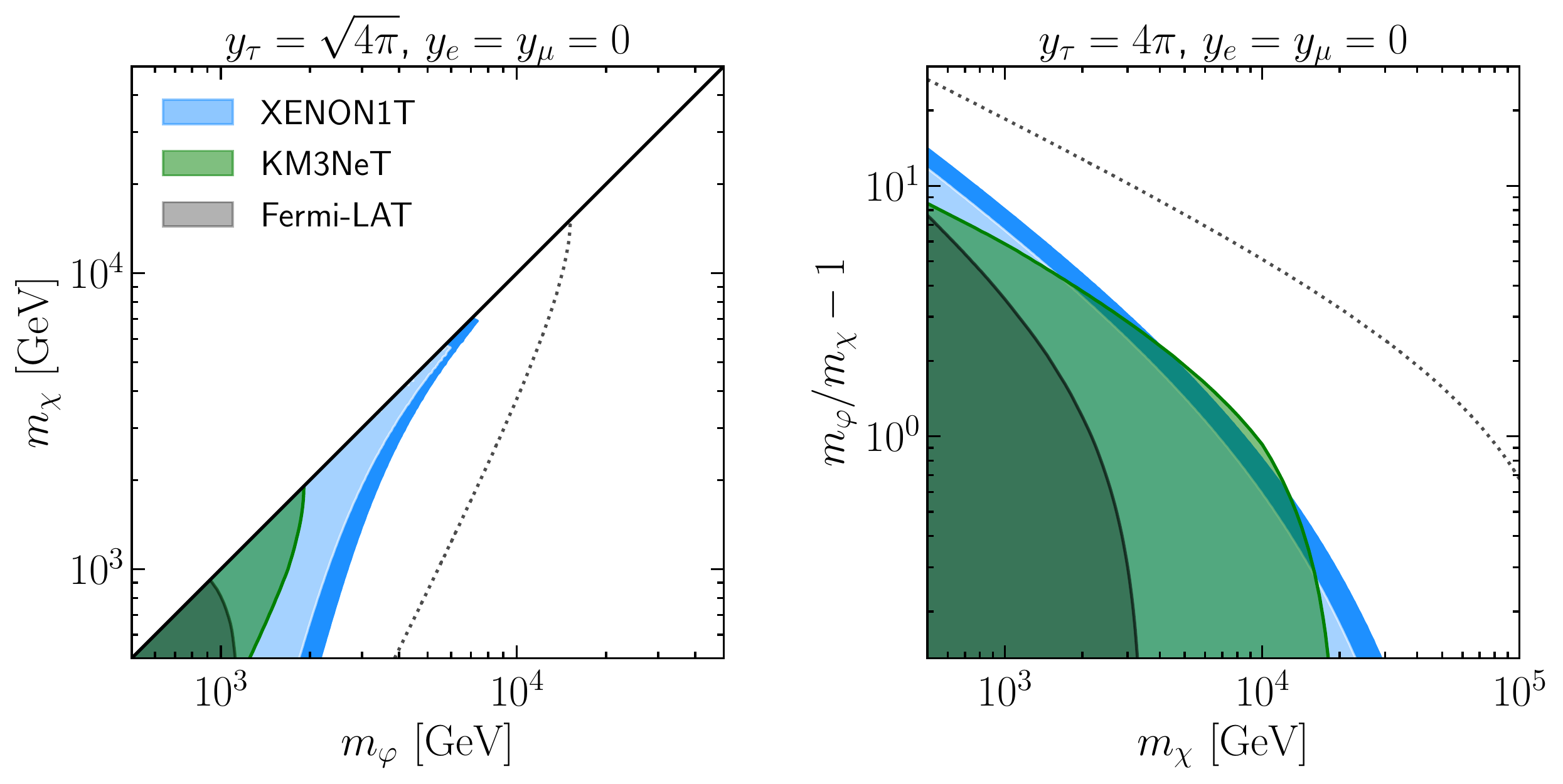}
	\caption{\textbf{Left:} KM3NeT projections (green shaded region) for the $t$-channel model compared with the existing Xenon1T (light blue shaded region) and Fermi-LAT (gray region) limits, as labelled, in the $\{ m_{\chi}, m_{\varphi} \}$ mass-plane. The blue band indicates the variation of the XENON1T 90\% CL upper limit with respect to the local density of dark matter, $\rho_{\odot}=(0.2, 0.4)\,\,\rm GeV / cm^3$. The coupling strength is fixed at $y_{\tau}=\sqrt{4 \pi}$. The dotted line denotes the relic density line. Inside the line the dark matter is under-abundant, while outside the dark matter is over-abundant. All exclusion limits and projection are provided at 90\% CL. \textbf{Right:} Same as left for a coupling strength fixed at $y_{\tau}=4\pi$ in the $\{ (m_\varphi/m_\chi-1),  m_{\chi} \}$-plane.}
	\label{fig:spin0}
\end{figure}

Figure~\ref{fig:spin0} summarises our findings in two projections of the dark matter and mediator mass planes for two different choices of $y_\tau$. It is clear that this simple scalar model is severely constrained by direct detection experiments. Even in the most optimistic scenario of very large coupling strength, the parameter space that could be potentially surveyed by future generation of neutrino telescopes is already being strongly disfavoured by XENON1T. The inclusion of the other lepton flavours does not change the picture, as it will have the effect of strengthening more the XENON1T with respect to the improve in the KM3NeT sensitivity. This can be understood as follows: the nature of annihilation process is $t$-channel, and therefore $\sigmav\propto 1/m_{\chi}^{2}$, which does not bode well for indirect detection searches. This is further supported by the small contribution given by the Fermi-LAT dSph constraints (gray shaded region). Fermi-LAT bounds occur of course because annihilation of dark matter particles into neutrinos produces neutrino lines, however annihilation into charged leptons unavoidably accompany this signal with equal branching ratio, because of the universality in the $y_\tau$ coupling. Moreover, as already said, neutrino final states  produce gamma rays because of electroweak corrections, contributing in slightly strengthening the Fermi-LAT bounds, as shown previously in~\cref{fig:Fermi_limit}. The blue band indicates the variation of the XENON1T 90\% CL upper limit with respect to the local density of dark matter: we see that smaller density imply obviously less stringent bounds, by a factor of $1/2$. This effect does not change our conclusion for this specific model, but illustrates the complementarity between dark matter searches, affected by different astrophysical uncertainties. 

Future improvements for $t$-channel models can be searched for in possibly a couple of directions. On the one side going beyond the simple picture of singlet dark matter will open up the parameter space. For instance triplet Dirac dark matter, described in~\cite{ElAisati:2017ppn}, leads to sizeable Sommerfeld enhancement~\cite{1931AnP...403..257S} because of the charged component in the triplet. This boosts $\sigmav$, while not affecting the direct detection bounds, favouring indirect detection searches. On the other hand, the inclusion into the model of a neutrino mass generation mechanism allows for a disentanglement between charged lepton and neutrino Yukawa-like couplings. Consequently exclusion bounds coming from charged leptons can be consistently weakened, allowing for more parameter space for future neutrino telescopes. This seems promising for instance for type II seesaw models~\cite{Hambye:2005tk,Arina:2011cu} or inverse seesaw (see ref.~\cite{CentellesChulia:2020dfh} for a recent review), where the neutrino Yukawa couplings are large by construction. A similar hierarchy could be applied to the case of $t$-channel models.

\subsection{Simplest Vector Mediator Models}\label{subsec:schan} 

Having in mind the two requirements described in our model building procedure, the most promising $s$-channel model features singlet Dirac dark matter candidate $\chi$ with a spin 1 mediator $Z'$.\footnote{This model is available in the \textsc{FeynRules}~\cite{Alloul:2013bka} database \url{https://feynrules.irmp.ucl.ac.be/wiki/DMsimp}, for usage see \eg~\cite{Andrea:2011ws,Backovic:2015soa,Mattelaer:2015haa,Neubert:2015fka}.} The interaction Lagrangian of the $Z'$ with the dark matter and the leptons is:
 \begin{align}\label{eq:schann}
 {\cal L}^{Z'} = \bar\chi \gamma_{\mu}
  (g^{V}_{\chi}+g^{A}_{\chi}\gamma_5)\chi\,Z'^{\mu}  + \sum_{\alpha} \bar l_\alpha \Big[\gamma_{\mu}
    (g^{V}_{l_{\alpha}}+g^{A}_{l_{\alpha}}\gamma_5) \Big] l_\alpha Z'^{\mu}\,,
\end{align}
where $l$ denotes leptons, ($\alpha=e, \mu,\tau$) is the flavour index, and $g^{V/A}$ are the vector/axial-vector couplings of dark matter and leptons with the $Z'$ respectively.  This model is very similar to the one used as benchmark at the LHC by the experimental collaborations~\cite{Aaboud:2019yqu,Sirunyan:2018exx}, except that $g^V_q=g^A_q=0$ in this case.

Concerning the mass generation mechanism for the $Z'$, the new boson with a vectorial coupling can get its mass either with the Stueckelberg mechanism~\cite{Stueckelberg:1900zz,Kors:2005uz} or via a new scalar boson (\ie a dark Higgs). In order to keep things minimal, we assume that if it exists, the new scalar is heavy enough that it can be integrated out or that it has very tiny couplings with the SM and dark matter particles not to induce additional interactions, hence the phenomenology of the dark matter presented in the following is not affected by the mediator mass generation mechanism. For building $Z'$ models we refer \eg to~\cite{Basso:2011na,Carena:2004xs}.

The desire to find dark matter models, which can be both optimally probed by KM3NeT and theoretically well motivated, will bring us to go beyond the purely simplified model approach and towards anomaly-free simplified models, presented in section~\ref{subsec:gaugedzp}. In order to get to our end goal, we investigate some interesting intermediary cases below where KM3NeT will exhibit good complementarity with other new physics searches. 

\paragraph{Pure vectorial coupling}
First let us consider the case where the coupling strengths in~\cref{eq:schann} are chosen as
\begin{equation}
    g^{V}_{l_{f}}\equiv g_l \,, \quad g^V_{\chi} \equiv g_\chi \quad {\rm and}\quad g^A_{l_{f}} = g^A_{\chi} = 0.
\end{equation}
\ie we have a purely vectorial coupling between dark matter and the leptonic sector couples equally to the $Z'$. The first choice is appealing to avoid issues with unitarity, see \ie~\cite{Kahlhoefer:2015bea}, but still requires additional particles to achieve anomaly freedom~\cite{Ellis:2017tkh}. The second choice is dictated by the fact that the mediator behaves like a gauge boson, hence it is reasonable to assume equal coupling for all flavours. The model is left with only four independent parameters, \ie two masses and two couplings: 
\begin{equation}
\{m_{\chi}, m_{Z'}, g_{\chi}, g_{l} \}\,.
\end{equation}
Additionally we consider equal leptonic and dark matter couplings, $g_{l}=g_\chi$. This is an arbitrary choice, but can be seen as an adequate benchmark for models where $g_{l} \sim g_\chi$. For this hierarchy of couplings, the diagram on the left in~\cref{fig:diagrams} dominates over the two mediator emission (central panel). Since the annihilation cross-section is proportional to both couplings in equal measure, $\sigmav \propto (g_{\chi}g_{l})^2$, indirect detection limits are insensitive to the particular hierarchy of couplings in this model.  The thermal annihilation cross-section for the $s$-channel process is of course maximised on resonance production, $2m_{\chi} \sim m_{Z'}$. The ultimate value on resonance is limited by the width of $Z'$ which is also set by the couplings $g_{\chi}$ and $g_{l}$.

We study the sensitivity reach of different experiments with respect to these parameters by either fixing or varying them accordingly which is shown in each figure. More specifically, we focus on dark matter and mediator masses above 500 GeV and up to 10 and 30 TeV respectively, well above the reach of LHC but exactly in the ball park of sensitivity for KM3NeT, as driven from the limits in the previous section. The couplings vary in between $10^{-1}$ and the perturbative upper value of $4\pi$. 

\begin{figure}
    \centering
    \includegraphics[width=\textwidth]{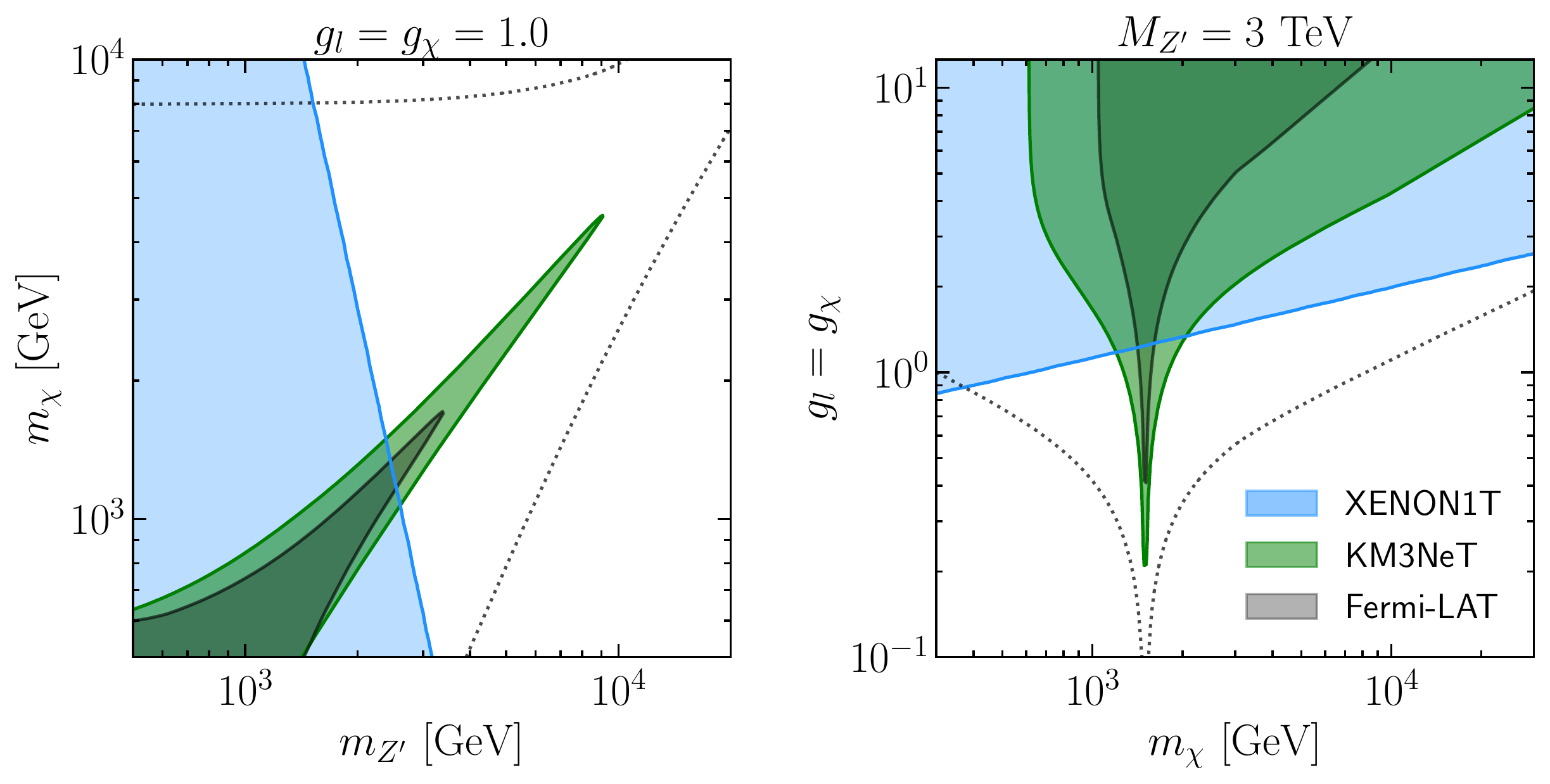}\\
    \includegraphics[width=\textwidth]{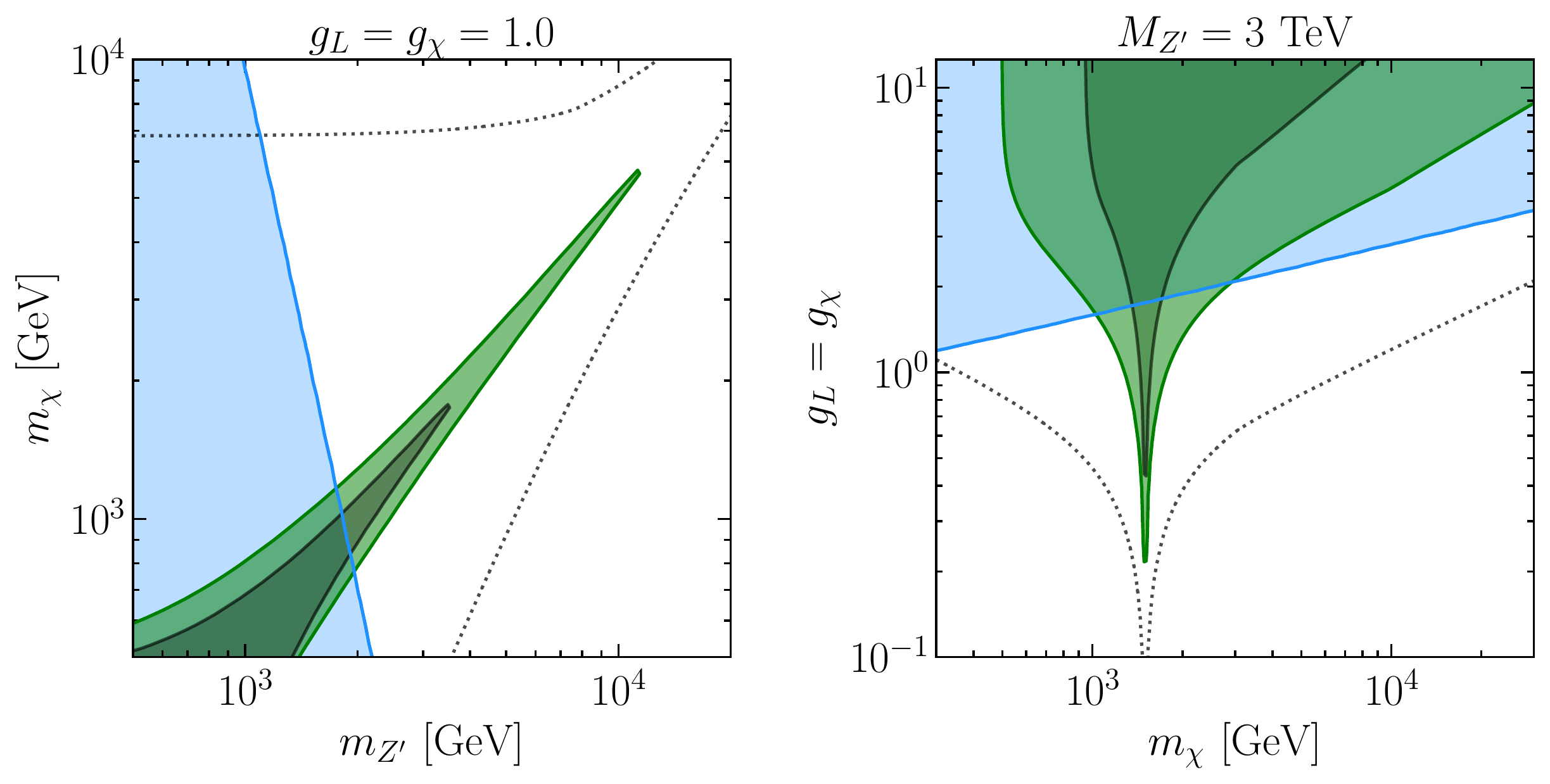}\\
    \caption{\textbf{Top Left:} The projected sensitivity of KM3NeT (green) together with current bounds from Fermi-LAT dSphs obtained with \textsc{MadDM} (gray)~\cite{Fermi-LAT:2016uux} and XENON1T (light blue)~\cite{Aprile:2018dbl} obtained with \textsc{RAPIDD}, as labelled, in the $\{m_{\chi}, m_{Z'} \}$-plane for fixed equal couplings of $g_{l,L}=g_{\chi}$. The black dotted lines are where $\sigmav= 2\times10^{-26} \, \rm cm^{3}/s$, \ie the value for simple freeze-out. All exclusion limits and projection are provided at 90\% CL. \textbf{Top Right:} Same as left in the $\{g_{l,L},m_\chi \}$-plane. 
    \textbf{Bottom row:} Same as top for the model with pure left-handed couplings.}
    \label{fig:spin1_simplest}
\end{figure}

Figure~\ref{fig:spin1_simplest} (upper panels) shows the parameter space for this model, varying the mediator and dark matter mass on the left plot in $\left(m_{\chi},m_{Z'}\right)$-plane where $g_{l}=g_{\chi}=1.0$ is taken, and varying coupling and dark matter mass on the right in comparison to other experimental searches. We see that the KM3NeT sensitivity for combined leptonic channels will be probing parameter spaces on the resonance which are currently beyond reach of Fermi-LAT. For sake of reference, we also show the thermal cross-section (black dotted line) that would give the correct relic density is around $\sim 2\times 10^{-26}$ cm$^{3}$/s.

Annihilation of dark matter particles into neutrinos will produce neutrino lines, however annihilation into leptons will accompany this signal with larger branching ratio. Indeed pure vectorial couplings penalise the neutrino channel due to the fact that there are no right-handed neutrinos in the SM. Therefore, above $m_{\tau}$, 
\begin{equation}\label{eq:pure_vec_branch}
    \sigmav_{\chi\chi\rightarrow \nu\overline{\nu}}= \frac{1}{2}\sigmav_{\chi\chi\rightarrow\ell^+\ell^-} = \frac{1}{3}\sigmav_{\rm{tot}}.
\end{equation}
A discrepancy in branching ratio may well be compensate by KM3NeT for $m_{\chi}>1$ TeV. However, since $\sigmav\propto1/m_{Z'}^4$, perturbative couplings can only achieve $\sigmav\sim10^{-24}\, \textrm{cm}^3/\textrm{s}$ when the mass differences of $m_{\chi}$ and $m_{Z'}$ are such that the annihilation is resonant. 
Notice that direct detection bounds are quite strong. For $s$-channel models the RGEs induce the usual pure vectorial spin-independent interactions with nuclei.
Within the time frame of operation of future indirect detection probes and neutrino telescopes, KM3NeT-ARCA will be competitive and add important additional information in the determination of the nature of dark matter. Notice however that this model is not the optimal target that KM3NeT will be able to probe, since the branching ratio into neutrinos is sub-dominant.

\paragraph{Pure left-handed coupling}
It is possible to construct a model that couples only to the $SU(2)_L$ lepton doublets of the SM. 
This is achieved by choosing the coupling configuration of~\cref{eq:schann} such that
\begin{equation}
 g^{V}_{l_{f}} =  g^{A}_{l_{f}}  \equiv \frac{1}{2} g_{\rm L}\,, \quad g^V_{\chi}\equiv g_\chi \quad {\rm and}\quad g^A_{\chi} = 0.
\end{equation}
This choice results again in four free parameters for the model:
\begin{equation}
\{m_{\chi}, m_{Z'}, g_{\chi}, g_{\rm L} \}\,,
\end{equation}
where we further set $g_\chi = g_L$ and perform the same scanning procedure as in the pure vectorial case. A pure left-handed coupling gives the more favourable scenario where 
\begin{equation}
    \sigmav_{\chi\chi\rightarrow \nu\overline{\nu}}= \sigmav_{\chi\chi\rightarrow\ell^+\ell^-} = \frac{1}{2}\sigmav_{\rm{tot}}.
\end{equation}
The effect of this choice is illustrated in~\cref{fig:spin1_simplest} (lowest panels), and affect mostly the contribution to dark matter-nucleon scattering for the case of the XENON1T experiment, which is weakened with respect to the case of pure vectorial couplings, carving out less of the viable parameter space for indirect searches. Nevertheless our conclusions are very similar to the case discussed previously. 

However with this model, it is arguable that we are on slightly less sturdy theoretical ground. Not only do we have a model that contains anomalies~\cite{Ellis:2017tkh}, but more immediately issues with unitarity violation occur, see \ie~\cite{Kahlhoefer:2015bea}. 

\subsection{Anomaly-free $L_{\mu}-L_{\tau}$ model}\label{subsec:gaugedzp}

A well  known, anomaly-free leptophilic scenario is the so-called $L_{\mu}-L_{\tau}$ gauge $U(1)$ model. This model assumes a specific gauge group under which the new particles are charged and gauged, which is defined as being the difference between muon- and tau-lepton numbers $L_\mu - L_\tau$. This model has been proposed in~\cite{He:1990pn,He:1991qd,Foot:1994vd,Baek:2001kca,Ma:2001md,Heeck:2011wj,Altmannshofer:2014cfa} for solving the long standing flavour anomalies and has been studied \ie in~\cite{Baek_2009,Biswas:2016yjr,Altmannshofer:2016jzy,Arcadi:2018tly} in connection with dark matter. 

With respect to~\cref{eq:schann}, the coupling configuration is set to:
\begin{equation}
 g^{V}_{\mu} = -g^{V}_{\tau}=g_{\mu-\tau}\,, \quad g^{V}_{e}=g^{A}_{l_{f}}=0\,, \quad g^V_\chi \equiv g_\chi \quad {\rm and}\quad g^A_{\chi} = 0.  
\end{equation}
All together the four free parameter of the model are:
\begin{equation}
\{m_{\chi}, m_{Z'}, g_{\chi}, g_{\mu-\tau} \}\,,
\end{equation}
which we first consider in the same range as in the previous section. 

\begin{figure}
    \centering
    \includegraphics[width=\textwidth]{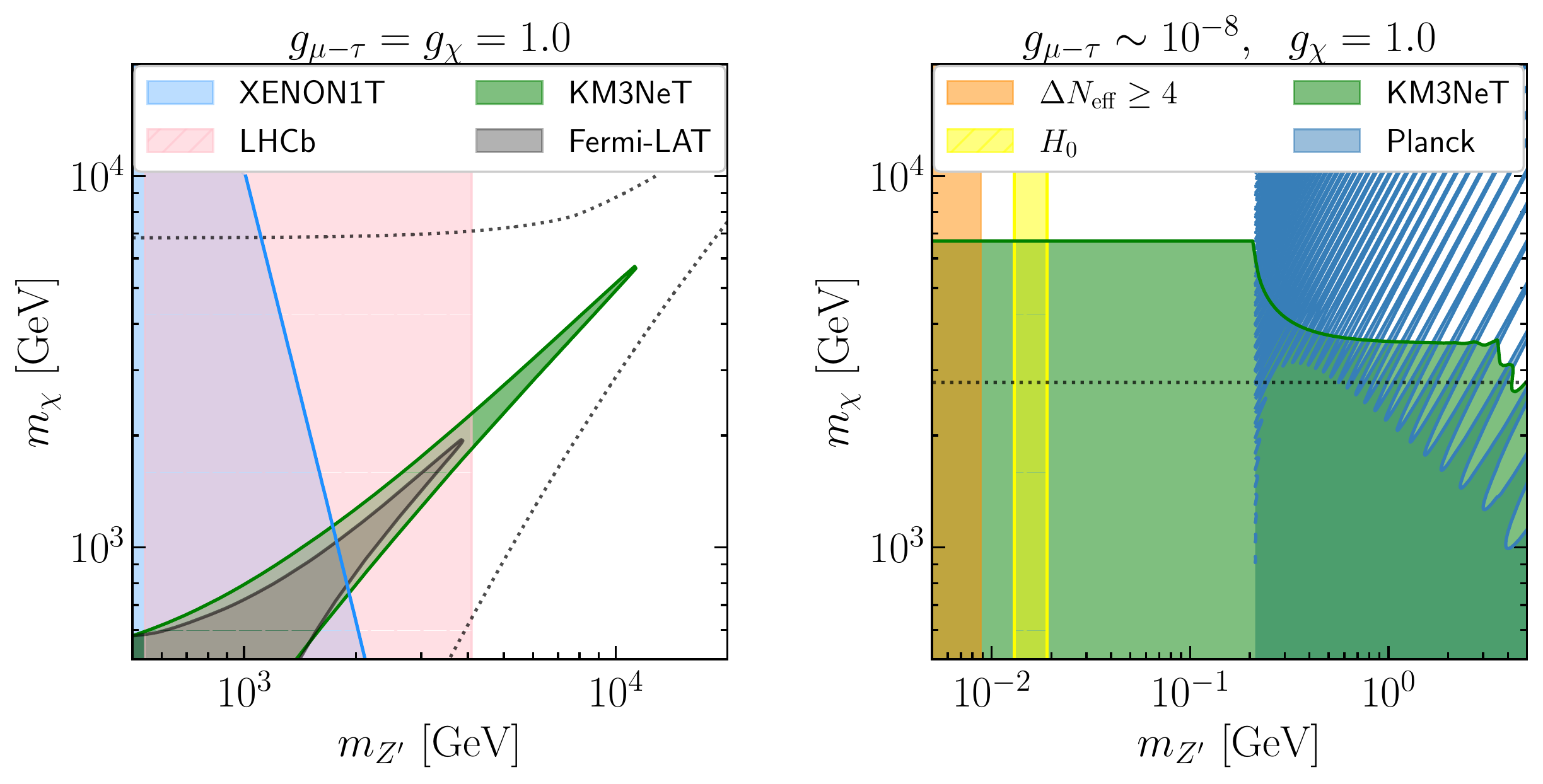}
    \caption{\textbf{Left:} Same as \cref{fig:spin1_simplest} for the anomaly-free $L_{\mu}-L_{\tau}$ model. We have added the favored region~\cite{Altmannshofer:2016jzy} for explaining the flavour anomalies found at LHCb~\cite{Aaij:2014ora,Aaij:2017vbb,Aaij:2019wad} (red shaded region). \textbf{Right:} The projected sensitivity of KM3NeT (green) together with current upper bound from Planck (blue)~\cite{Bringmann:2016din,Slatyer:2015jla,ArkaniHamed:2008qn}, as labelled, in the $\{m_{\chi}, m_{Z'} \}$-plane for fixed  couplings of $g_{\mu-\tau} << g_{\chi} = 1$. The yellow region denotes the model parameter space favoured to alleviate the Hubble tension~\cite{Bernal:2016gxb,Escudero:2019gzq}. The dotted black line represents the relic density line.} 
    \label{fig:spin1_LmuLtau}
\end{figure}

The expected sensitivity for the KM3NeT as well as the constraints from different experiments using this model are shown in~\cref{fig:spin1_LmuLtau} left, when $g_{\mu-\tau}=g_{\chi}=1$ is satisfied. 
Besides having a well motivated theoretical model, we are in the same position as the simple $Z'$ models concerning the complementarity among direct, indirect dark matter bounds and the KM3NeT sensitivity.  Despite the diminished branching ratio into neutrinos, the fact that in this model, there is no penalty for not considering the electron channel, the KM3NeT projection does not weaken as much as in~\cref{fig:spin1_simplest}, hence the results are more similar to the left-handed case than the pure vector one.
Two other important comments are in order.
First, the direct detection limits are not due to RGEs, as the $U(1)_{\mu-\tau}$ has the property that the RGE effects completely cancel. Therefore, the major contribution comes from the kinetic mixing $\varepsilon$, which will be generated at the loop level. The loop induced kinetic mixing is finite~\cite{Araki:2017wyg}
\begin{equation}
\varepsilon\left(q^{2}\right)= \frac{e g_{\mu-\tau}}{12 \pi^{2}} \ln \frac{m_{\mu}^{2}}{m_{\tau}^{2}},
\end{equation}
and leads to sizeable values, for instance $\varepsilon\sim 10^{-2}$ for $g_{\mu-\tau} \simeq 1$. 
Secondly, the same model can accommodate the favored region for flavor anomalies observed at LHCb~\cite{Aaij:2014ora,Aaij:2017vbb,Aaij:2019wad}, red shaded region in the plot of~\cref{fig:spin1_LmuLtau}. In order for this model to account for anomalies in $b \to s \mu^+ \mu^-$ decays, it is necessary to assume that there is additional new physics. For our scope, a fourth generation of vector-like quarks coupling to the $Z'$ is a viable solution, as these new particles can be taken much heavier than the dark matter scale we are interested of, so that they can be safely integrated out without affecting our phenomenological predictions~\cite{Alguero:2019ptt,Altmannshofer:2019xda,DiLuzio:2019jyq}. For details about the favoured region derivation see \ie~\cite{Altmannshofer:2016jzy,Altmannshofer:2015mqa,Altmannshofer:2014cfa}.
\\
From~\cref{fig:spin1_LmuLtau} this gauged model has the benefit of being gauge invariant but does not seem to give KM3NeT a huge competitive edge. However, this model can easily accommodate a light $Z'$, which is exactly the same ball park of mediator masses employed to explain the $(g-2)_{\mu}$ anomaly~\cite{Bennett:2002jb,Bennett:2004pv,Bennett:2006fi,Roberts:2010cj}, see \eg the recent ref.~\cite{Amaral:2020tga}. We use the fact that, for $m_{Z'}< 2\, m_{\mu}$, the branching ratio to neutrinos is 100\%. This is favourable to neutrino telescopes, both those such as Super-Kamiokande~\cite{Abe:2020sbr}, which are sensitive to light dark matter~\cite{Klop:2018ltd, Asai:2020qlp}, as well as those most sensitive to heavier dark matter, such as KM3NeT.

Of course, with a new light vector, relatively large couplings with SM particles are strongly constrained by colliders and precision experiments. 
It is, however, completely legitimate to take $g_{\mu-\tau}\ll g_{\chi}$. This model comes under the category of secluded dark matter models~\cite{Pospelov:2007mp}, and specifically features a very light mediator ($m_{Z'} \ll m_\chi$), large and perturbative $g_\chi$ while negligible coupling to the SM sector, $g_{\mu-\tau} \lesssim 10^{-4}$ at least. This last requirement is set to evade strong constraints on light fields coupling to SM leptons, but the meaning of this model is deeper than simply avoiding experimental bounds. For such coupling strength hierarchy, the dominant annihilation channel is $\chi \bar{\chi} \to Z' Z'$ (see~\cref{fig:diagrams} centre), which implies that the dark matter can achieve the correct relic density independently of its coupling with the SM, because $\sigmav$ depends only on $g_\chi^4$. Since the $Z'$ is not stable, it will eventually decay into four SM neutrinos, even though $g_{\mu-\tau}$ is extremely tiny. Depending on the $m_{Z'}$ value, annihilation into charged leptons, when kinematically allowed, takes place with equal branching ratios and gives rise to a $4\mu$ or $4\tau$ final state. For simplicity we do not consider mixed final states in this study. 
Very light $Z'$ mediators are boosted in the very heavy dark matter annihilation reference frame. Subsequently, they decay into neutrinos producing characteristic box-shaped neutrino signals. 
While the box-shaped gamma-ray signals have already been explored in certain depth in the literature, see \ie~\cite{Ibarra:2012dw,Ibarra:2013eda,Ibarra:2015tya,Leane:2017vag,Arina:2017sng}, box-shaped neutrino signals have only been poorly studied, see~\cite{ElAisati:2017ppn}.

Let us now quantify how constrained is the model parameter space in the light of current bounds. In the secluded regime, high intensity experiments constrain $g_{\mu-\tau}\lesssim 10^{-4}$ for $m_{Z'}\lesssim 10$ GeV~\cite{Escudero:2019gzq}, while Big Bang Nucleosynthesis (BBN) sets a lower bound on the coupling strength not to spoil its accurate predictions in terms of element abundances with the injection of energy. Indeed the $\chi\chi\rightarrow Z'Z'$ process is independent to $g_{\mu-\tau}$, however one cannot choose an arbitrarily small value as a sufficiently long lived $Z'$ will disrupt BBN. A conservative bound would be taking the lifetime sufficiently short such as
\begin{equation}
\tau_{Z'} = \sum_{f}\frac{12\pi m_{Z'}^2}{g_{\mu-\tau}^2 \,\sqrt{m_{Z'}^2 - 4m_f^2}\left(m_{Z'}^2 -m_f^2\right)}\lesssim 1 \rm \,  s,
\end{equation}
where $m_f$ is the mass of the products of decay and $f$ runs over the fermionic decays kinematically allowed. This bound sets $g_{\mu-\tau} > 10^{-10}$ for $m_{Z'}=1$ MeV, while being looser at larger dark matter masses. Notice that scenarios for the $(g-2)_{\mu}$ anomaly favour couplings of $g_{\mu-\tau}\sim10^{-5}$. 
A more stringent bound comes from measurements of $\Delta N_{\rm{eff}}$, the effective relativistic degree of freedom before recombination~\cite{Escudero:2019gzq,Aghanim:2018eyx}, which sets a bound on the $Z'$ mass to be larger than 4 MeV roughly, by asking $\Delta N_{\rm eff} \leq 4$.
The constraints on a light $Z'$ are of course independent of dark matter, so we have to add direct dark matter detection limits as before. Now however, the smaller values of $m_{Z'}$ do not suppress the recoil rate so much, resulting in strong constraints on $g_{\mu-\tau}$, which are relevant for $m_\chi > 6$ GeV and set $g_{\mu-\tau} < 10^{-6}$ from XENON1T. We have checked that  $g_{\mu-\tau} < 10^{-8}$ produces nuclear recoil event rates well below  the sensitivity of future direct detection experiments. We see that in between the upper and lower bound there is still room to play safe and evade all constraints. What can not be avoided are indirect detection bounds, which come from Fermi-LAT and Planck~\cite{Aghanim:2018eyx}. Indeed $\sigmav$ is severely boosted by the small velocities in dSphs or at the recombination epoch, as a light mediator induces Sommerfeld enhancement, which we properly include as in~\cite{Hisano:2004ds,ArkaniHamed:2008qn,Iengo:2009ni,Cassel:2009wt,Arina:2010wv}. Lastly, self-interaction constrains the size of the scattering process $\chi \chi \to \chi \chi$, see \eg~\cite{Bringmann:2016din}, impacting the region of small dark matter masses and very light mediators. 

The results are shown in the right panel of~\cref{fig:spin1_LmuLtau}. We see for  $m_{Z'} > 2 m_\mu$ the model parameter space is strongly disfavoured by current bound from Planck, computed as in~\cite{Bringmann:2016din,Slatyer:2015jla,ArkaniHamed:2008qn}. This bound supersedes Fermi-LAT dSph upper limit (not shown) because dSphs are warmer than the epoch of recombination, and this is reflected in a smaller Sommerfeld boost.  However neutrino telescopes can access the model parameter space below the kinematic threshold of the muon final state, being able to probe the model parameter space proving the correct relic density. The dip in the KM3NeT bound is due to the opening of the charged lepton final states, which increases $\sigmav$.
In the same ball park, $m_{Z'}\sim 10 $ MeV can also contribute to the resolution of the Hubble tension~\cite{Aghanim:2018eyx,Riess:2016jrr,Riess:2018byc,Riess:2019cxk} (yellow shaded band). More specifically the light vector would contribute to increase $\Delta N_{\rm eff}$ up to roughly $0.2-0.4$~\cite{Bernal:2016gxb,Escudero:2019gzq}, value that can reconcile the determination of $H_0$ from local measurements with the one from the cosmic microwave background.
Self-interaction bounds are not shown to avoid cluttering, as they exclude a small portion of the lower left corner, which is already disfavoured by $\Delta N_{\rm eff}$.
Lastly, notice that we did not include electroweak radiation from the prompt neutrino final states. This would have the effect of generating a bound from the recombination epoch for dark matter above the TeV, where electroweak corrections are sizeable.
All in all the region below the muon threshold is a sweet spot for revealing the best capabilities of KM3NeT, especially for dark matter below the TeV scale, but it remains to future work to determine whether electroweak corrections provide Planck with the means to constrain this region. 

\section{Conclusions} \label{sec:conclusion}
 
In this paper we have discussed the future prospects of a KM3NeT-like neutrino telescope in terms of the expected sensitivity for annihilating dark matter. The detector properties and location are particularly favourable to search for such signals from the centre of the Milky Way.  We have embarked on a global investigation, assessing how neutrino telescopes will contribute to the broader search for WIMP-like dark matter particles. We have explored the most promising, minimal scenarios where neutrino telescopes will play a dominant role. Some of these scenarios can be associated with existing hints of new physics found in particle physics and cosmology.

Dark matter particles are expected to correlate with the galactic center, imprinting an anisotropic feature on the neutrino sky.
Through an angular power spectrum analysis, we have assessed the angular distributions of neutrinos to probe dark matter signals on top of the atmospheric background. We have assumed a 10-year running period once the experiment is fully operating. We have simulated neutrino skymaps with through-going track events, taking into account the contribution of both extragalactic and galactic dark matter annihilations. Using a Monte Carlo method, we have set model-independent upper bounds on the thermally averaged dark matter annihilation cross-section and dark matter mass, with a particular focus on leptonic channels which provide the best projections. More specifically, we have considered a dark matter mass range of 200 -- $10^5$ GeV, ideally probed by the KM3NeT-ARCA configuration.  

Moreover, we have fully incorporated the electroweak corrections from the charged lepton final states and neutrino lines, that significantly modify the neutrino and gamma-ray energy spectra. We have additionally encompassed the effect of the large systematic uncertainties arising from the distribution of the dark matter density profile, however we did not include experimental uncertainties. As demonstrated, the angular power spectrum analysis is robust against assumptions of the galactic dark matter distribution in this mass range. Our limits overall weaken only by $\sim40\%$ when considering the Burkert dark matter density profile as opposed to NFW. Although, the obtained limits are suggestive in terms of the potential reach of the experiment, it should be noted that the searches can be improved and extended to the lower dark matter mass range by the addition of KM3NeT-ORCA site which will change the  bounds in~\cref{fig:Fermi_limit} for example. Employing different analysis and event reconstruction techniques that are not incorporated in this paper but subject to ongoing studies, the results can be improved further. For example, on top of the tracks from charge current interactions, addition of different types of event topologies (tracks and shower events) could possibly enhance the obtained bounds.    

We have interpreted the projected sensitivities with respect to minimal extensions of the SM that include a Dirac dark matter candidate and a mediator. The most promising for neutrino telescopes couples dark matter to SM leptons only. Even still, neutrino line signals usually occur in conjunction with charged lepton emission.
We have shown that the latter provides significant upper bounds from present gamma-ray telescopes. Additionally, renormalization and loop effects provoke interactions with quarks and gluons. This means direct dark matter detection experiments are also relevant. 
For example, XENON1T, almost completely dominates over the parameter space for the $t$-channel model, whereas it leaves room for discovery/exploration for the future KM3NeT-ARCA for the $s$-channel model.  Whilst the minimal $s$-channel model we have considered suffers from theoretical issues, we have further investigated an anomaly free and gauge invariant scenario, the so-called $L_{\mu} -L_{\tau}$ model. This produces a phenomenology analogous to our minimal setup when mediator masses are heavy. It is remarkably interesting for neutrino telescopes in its secluded realization, featuring a very light mediator mass below the muon mass threshold. Indeed in this specific mass range, the model can ultimately be surveyed by neutrino telescopes, which appear to be sensitive to thermal values of $\sigmav$, as demonstrated in~\cref{fig:spin1_LmuLtau}. 

There are several possibilities to further improve the impact of upcoming neutrino telescopes in the search for dark matter signals. The minimal choice of models can be broadened, to encompass a wider portion of the viable theoretical landscape. For instance the study of pure neutrino or anti-neutrino lines ($\nu_l \nu_l$ or $\bar{\nu}_l \bar{\nu}_l$) is an interesting possibility as those signals do not have a charged fermionic counter part. Alternatively, the inclusion of non-minimal dark matter multiplets, see~\cite{ElAisati:2017ppn}, would provide additional effects such as Sommerfeld enhancement, increasing the sensitivity of indirect detection bounds, while leaving unaltered the reach of direct detection searches. 

By presenting our projections in the context of multiple experimental searches, we demonstrate a high degree of complementarity. As always, gamma-ray telescopes and direct dark matter detection experiments play an important role, but Neutrino telescopes will be able to probe new regions of parameter space. We stress that independent experimental probes are vital for comprehensive search for dark matter. Each experiment has its own set-up and sources of uncertainty. A diverse range of probes is necessary to reduce bias, minimizing the impact of individual uncertainties and clarifies the particle physics interpretation. 

Improvements both in the analysis methods and theoretical implications of upcoming neutrino telescopes will have a large impact on our understanding of dark matter physics.
They will additionally provide guidelines on how KM3NeT-like telescopes will complement other dark matter searches in the next decade.

\acknowledgments
We thank K. Ng for enlightening discussions and D. Massaro for help with the MadDM code. We also acknowledge A. Heijboer and R. Muller for clarification of the KM3NeT-ARCA specifications, B. Kavanagh for fixing a bug in \textsc{RunDM}, J. Heisig, M. Hostert and M. Kirk for discussions. SB is supported by NWO (Dutch National Science Organisation) under project no: 680-91-004. CA is supported by the Innoviris  ATTRACT 2018 104 BECAP 2 agreement. AC is supported by the F.R.S.-FNRS under the Excellence of Science EOS be.h project n. 30820817. MC acknowledges partial support from the research grant number 2017W4HA7S “NAT-NET: Neutrino and Astroparticle Theory Network” under the program PRIN 2017 funded by the Italian Ministero dell’Università e della Ricerca (MUR) and from the research project TAsP (Theoretical Astroparticle Physics) funded by the Istituto Nazionale di Fisica Nucleare (INFN).
SA acknowledges the support by JSPS/MEXT KAKENHI Grant Numbers JP17H04836, JP18H04340, JP20H05850, and JP20H05861 (SA).

\bibliographystyle{jhep}
\bibliography{references}

\providecommand{\href}[2]{#2}\begingroup\raggedright\begin{thebibliography}{100}

\bibitem{Arguelles:2019ouk}
C.~A. Arg\"uelles, A.~Diaz, A.~Kheirandish, A.~Olivares-Del-Campo, I.~Safa and
  A.~C. Vincent, \emph{{Dark Matter Annihilation to Neutrinos}},
  \href{https://arxiv.org/abs/1912.09486}{{\ttfamily 1912.09486}}.

\bibitem{Baur:2019jwm}
{\scshape IceCube} collaboration, \emph{{Dark matter searches with the IceCube
  Upgrade}}, \href{https://doi.org/10.22323/1.358.0506}{\emph{PoS} {\bfseries
  ICRC2019} (2020) 506} [\href{https://arxiv.org/abs/1908.08236}{{\ttfamily
  1908.08236}}].

\bibitem{Aartsen:2019swn}
{\scshape IceCube} collaboration, \emph{{Neutrino astronomy with the next
  generation IceCube Neutrino Observatory}},
  \href{https://arxiv.org/abs/1911.02561}{{\ttfamily 1911.02561}}.

\bibitem{Avrorin:2014vca}
A.~D. Avrorin et~al., \emph{{Sensitivity of the Baikal-GVD neutrino telescope
  to neutrino emission toward the center of the galactic dark matter halo}},
  \href{https://doi.org/10.1134/S0021364015050021}{\emph{JETP Lett.} {\bfseries
  101} (2015) 289} [\href{https://arxiv.org/abs/1412.3672}{{\ttfamily
  1412.3672}}].

\bibitem{Agostini:2020aar}
{\scshape P-ONE} collaboration, \emph{{The Pacific Ocean Neutrino Experiment}},
  \href{https://doi.org/10.1038/s41550-020-1182-4}{\emph{Nature Astron.}
  {\bfseries 4} (2020) 913} [\href{https://arxiv.org/abs/2005.09493}{{\ttfamily
  2005.09493}}].

\bibitem{Adri_n_Mart_nez_2016}
S.~Adri{\'{a}}n-Mart{\'{\i}}nez, M.~Ageron, F.~Aharonian, S.~Aiello, A.~Albert,
  F.~Ameli et~al., \emph{Letter of intent for {KM}3net 2.0},
  \href{https://doi.org/10.1088/0954-3899/43/8/084001}{\emph{Journal of Physics
  G: Nuclear and Particle Physics} {\bfseries 43} (2016) 084001}.

\bibitem{Benito:2019ngh}
M.~Benito, A.~Cuoco and F.~Iocco, \emph{{Handling the Uncertainties in the
  Galactic Dark Matter Distribution for Particle Dark Matter Searches}},
  \href{https://doi.org/10.1088/1475-7516/2019/03/033}{\emph{JCAP} {\bfseries
  03} (2019) 033} [\href{https://arxiv.org/abs/1901.02460}{{\ttfamily
  1901.02460}}].

\bibitem{Benito:2020lgu}
M.~Benito, F.~Iocco and A.~Cuoco, \emph{{Uncertainties in the Galactic dark
  matter distribution: an update}},
  \href{https://arxiv.org/abs/2009.13523}{{\ttfamily 2009.13523}}.

\bibitem{Abbasi:2011eq}
{\scshape IceCube} collaboration, \emph{{Search for dark matter from the
  Galactic halo with the IceCube Neutrino Telescope}},
  \href{https://doi.org/10.1103/PhysRevD.84.022004}{\emph{Phys. Rev. D}
  {\bfseries 84} (2011) 022004}
  [\href{https://arxiv.org/abs/1101.3349}{{\ttfamily 1101.3349}}].

\bibitem{Adrian-Martinez:2015wey}
{\scshape ANTARES} collaboration, \emph{{Search of Dark Matter Annihilation in
  the Galactic Centre using the ANTARES Neutrino Telescope}},
  \href{https://doi.org/10.1088/1475-7516/2015/10/068}{\emph{JCAP} {\bfseries
  10} (2015) 068} [\href{https://arxiv.org/abs/1505.04866}{{\ttfamily
  1505.04866}}].

\bibitem{Aartsen:2017ulx}
{\scshape IceCube} collaboration, \emph{{Search for Neutrinos from Dark Matter
  Self-Annihilations in the center of the Milky Way with 3 years of
  IceCube/DeepCore}},
  \href{https://doi.org/10.1140/epjc/s10052-017-5213-y}{\emph{Eur. Phys. J. C}
  {\bfseries 77} (2017) 627}
  [\href{https://arxiv.org/abs/1705.08103}{{\ttfamily 1705.08103}}].

\bibitem{Albert:2016emp}
A.~Albert et~al., \emph{{Results from the search for dark matter in the Milky
  Way with 9 years of data of the ANTARES neutrino telescope}},
  \href{https://doi.org/10.1016/j.physletb.2017.03.063}{\emph{Phys. Lett. B}
  {\bfseries 769} (2017) 249}
  [\href{https://arxiv.org/abs/1612.04595}{{\ttfamily 1612.04595}}].

\bibitem{ANTARES:2019svn}
{\scshape ANTARES} collaboration, \emph{{Search for dark matter towards the
  Galactic Centre with 11 years of ANTARES data}},
  \href{https://doi.org/10.1016/j.physletb.2020.135439}{\emph{Phys. Lett. B}
  {\bfseries 805} (2020) 135439}
  [\href{https://arxiv.org/abs/1912.05296}{{\ttfamily 1912.05296}}].

\bibitem{Aartsen:2020tdl}
{\scshape ANTARES, IceCube} collaboration, \emph{{Combined search for neutrinos
  from dark matter self-annihilation in the Galactic Center with ANTARES and
  IceCube}}, \href{https://doi.org/10.1103/PhysRevD.102.082002}{\emph{Phys.
  Rev. D} {\bfseries 102} (2020) 082002}
  [\href{https://arxiv.org/abs/2003.06614}{{\ttfamily 2003.06614}}].

\bibitem{Benito:2016kyp}
M.~Benito, N.~Bernal, N.~Bozorgnia, F.~Calore and F.~Iocco, \emph{{Particle
  Dark Matter Constraints: the Effect of Galactic Uncertainties}},
  \href{https://doi.org/10.1088/1475-7516/2017/02/007}{\emph{JCAP} {\bfseries
  02} (2017) 007} [\href{https://arxiv.org/abs/1612.02010}{{\ttfamily
  1612.02010}}].

\bibitem{Aartsen:2014hva}
{\scshape IceCube} collaboration, \emph{{Multipole analysis of IceCube data to
  search for dark matter accumulated in the Galactic halo}},
  \href{https://doi.org/10.1140/epjc/s10052-014-3224-5}{\emph{Eur. Phys. J. C}
  {\bfseries 75} (2015) 20} [\href{https://arxiv.org/abs/1406.6868}{{\ttfamily
  1406.6868}}].

\bibitem{Dekker:2019gpe}
A.~Dekker, M.~Chianese and S.~Ando, \emph{{Probing dark matter signals in
  neutrino telescopes through angular power spectrum}},
  \href{https://doi.org/10.1088/1475-7516/2020/09/007}{\emph{JCAP} {\bfseries
  09} (2020) 007} [\href{https://arxiv.org/abs/1910.12917}{{\ttfamily
  1910.12917}}].

\bibitem{Fermi-LAT:2016uux}
{\scshape DES, Fermi-LAT} collaboration, \emph{{Searching for Dark Matter
  Annihilation in Recently Discovered Milky Way Satellites with Fermi-LAT}},
  \href{https://doi.org/10.3847/1538-4357/834/2/110}{\emph{Astrophys. J.}
  {\bfseries 834} (2017) 110}
  [\href{https://arxiv.org/abs/1611.03184}{{\ttfamily 1611.03184}}].

\bibitem{Lindner_2010}
M.~Lindner, A.~Merle and V.~Niro, \emph{Enhancing dark matter annihilation into
  neutrinos}, \href{https://doi.org/10.1103/physrevd.82.123529}{\emph{Physical
  Review D} {\bfseries 82} (2010) }.

\bibitem{ElAisati:2017ppn}
C.~El~Aisati, C.~Garcia-Cely, T.~Hambye and L.~Vanderheyden, \emph{{Prospects
  for discovering a neutrino line induced by dark matter annihilation}},
  \href{https://doi.org/10.1088/1475-7516/2017/10/021}{\emph{JCAP} {\bfseries
  10} (2017) 021} [\href{https://arxiv.org/abs/1706.06600}{{\ttfamily
  1706.06600}}].

\bibitem{Kahlhoefer:2015bea}
F.~Kahlhoefer, K.~Schmidt-Hoberg, T.~Schwetz and S.~Vogl, \emph{{Implications
  of unitarity and gauge invariance for simplified dark matter models}},
  \href{https://doi.org/10.1007/JHEP02(2016)016}{\emph{JHEP} {\bfseries 02}
  (2016) 016} [\href{https://arxiv.org/abs/1510.02110}{{\ttfamily
  1510.02110}}].

\bibitem{Ellis:2017tkh}
J.~Ellis, M.~Fairbairn and P.~Tunney, \emph{{Anomaly-Free Dark Matter Models
  are not so Simple}},
  \href{https://doi.org/10.1007/JHEP08(2017)053}{\emph{JHEP} {\bfseries 08}
  (2017) 053} [\href{https://arxiv.org/abs/1704.03850}{{\ttfamily
  1704.03850}}].

\bibitem{He:1990pn}
X.~He, G.~C. Joshi, H.~Lew and R.~Volkas, \emph{{NEW Z-prime PHENOMENOLOGY}},
  \href{https://doi.org/10.1103/PhysRevD.43.R22}{\emph{Phys. Rev. D} {\bfseries
  43} (1991) 22}.

\bibitem{He:1991qd}
X.-G. He, G.~C. Joshi, H.~Lew and R.~Volkas, \emph{{Simplest Z-prime model}},
  \href{https://doi.org/10.1103/PhysRevD.44.2118}{\emph{Phys. Rev. D}
  {\bfseries 44} (1991) 2118}.

\bibitem{Foot:1994vd}
R.~Foot, X.~He, H.~Lew and R.~Volkas, \emph{{Model for a light Z-prime boson}},
  \href{https://doi.org/10.1103/PhysRevD.50.4571}{\emph{Phys. Rev. D}
  {\bfseries 50} (1994) 4571}
  [\href{https://arxiv.org/abs/hep-ph/9401250}{{\ttfamily hep-ph/9401250}}].

\bibitem{Baek:2001kca}
S.~Baek, N.~Deshpande, X.~He and P.~Ko, \emph{{Muon anomalous g-2 and gauged
  L(muon) - L(tau) models}},
  \href{https://doi.org/10.1103/PhysRevD.64.055006}{\emph{Phys. Rev. D}
  {\bfseries 64} (2001) 055006}
  [\href{https://arxiv.org/abs/hep-ph/0104141}{{\ttfamily hep-ph/0104141}}].

\bibitem{Ma:2001md}
E.~Ma, D.~Roy and S.~Roy, \emph{{Gauged L(mu) - L(tau) with large muon
  anomalous magnetic moment and the bimaximal mixing of neutrinos}},
  \href{https://doi.org/10.1016/S0370-2693(01)01428-9}{\emph{Phys. Lett. B}
  {\bfseries 525} (2002) 101}
  [\href{https://arxiv.org/abs/hep-ph/0110146}{{\ttfamily hep-ph/0110146}}].

\bibitem{Heeck:2011wj}
J.~Heeck and W.~Rodejohann, \emph{{Gauged L\_mu - L\_tau Symmetry at the
  Electroweak Scale}},
  \href{https://doi.org/10.1103/PhysRevD.84.075007}{\emph{Phys. Rev. D}
  {\bfseries 84} (2011) 075007}
  [\href{https://arxiv.org/abs/1107.5238}{{\ttfamily 1107.5238}}].

\bibitem{Altmannshofer:2014cfa}
W.~Altmannshofer, S.~Gori, M.~Pospelov and I.~Yavin, \emph{{Quark flavor
  transitions in $L_\mu-L_\tau$ models}},
  \href{https://doi.org/10.1103/PhysRevD.89.095033}{\emph{Phys. Rev. D}
  {\bfseries 89} (2014) 095033}
  [\href{https://arxiv.org/abs/1403.1269}{{\ttfamily 1403.1269}}].

\bibitem{Catena:2009mf}
R.~Catena and P.~Ullio, \emph{{A novel determination of the local dark matter
  density}}, \href{https://doi.org/10.1088/1475-7516/2010/08/004}{\emph{JCAP}
  {\bfseries 08} (2010) 004} [\href{https://arxiv.org/abs/0907.0018}{{\ttfamily
  0907.0018}}].

\bibitem{Pato:2015dua}
M.~Pato, F.~Iocco and G.~Bertone, \emph{{Dynamical constraints on the dark
  matter distribution in the Milky Way}},
  \href{https://doi.org/10.1088/1475-7516/2015/12/001}{\emph{JCAP} {\bfseries
  12} (2015) 001} [\href{https://arxiv.org/abs/1504.06324}{{\ttfamily
  1504.06324}}].

\bibitem{Nesti:2013uwa}
F.~Nesti and P.~Salucci, \emph{{The Dark Matter halo of the Milky Way, AD
  2013}}, \href{https://doi.org/10.1088/1475-7516/2013/07/016}{\emph{JCAP}
  {\bfseries 07} (2013) 016} [\href{https://arxiv.org/abs/1304.5127}{{\ttfamily
  1304.5127}}].

\bibitem{Cirelli:2010xx}
M.~Cirelli, G.~Corcella, A.~Hektor, G.~Hutsi, M.~Kadastik, P.~Panci et~al.,
  \emph{{PPPC 4 DM ID: A Poor Particle Physicist Cookbook for Dark Matter
  Indirect Detection}},
  \href{https://doi.org/10.1088/1475-7516/2012/10/E01}{\emph{JCAP} {\bfseries
  03} (2011) 051} [\href{https://arxiv.org/abs/1012.4515}{{\ttfamily
  1012.4515}}].

\bibitem{Ciafaloni:2010ti}
P.~Ciafaloni, D.~Comelli, A.~Riotto, F.~Sala, A.~Strumia and A.~Urbano,
  \emph{{Weak Corrections are Relevant for Dark Matter Indirect Detection}},
  \href{https://doi.org/10.1088/1475-7516/2011/03/019}{\emph{JCAP} {\bfseries
  03} (2011) 019} [\href{https://arxiv.org/abs/1009.0224}{{\ttfamily
  1009.0224}}].

\bibitem{Bauer:2020jay}
C.~W. Bauer, N.~L. Rodd and B.~R. Webber, \emph{{Dark Matter Spectra from the
  Electroweak to the Planck Scale}},
  \href{https://arxiv.org/abs/2007.15001}{{\ttfamily 2007.15001}}.

\bibitem{Ibarra:2012dw}
A.~Ibarra, S.~Lopez~Gehler and M.~Pato, \emph{{Dark matter constraints from
  box-shaped gamma-ray features}},
  \href{https://doi.org/10.1088/1475-7516/2012/07/043}{\emph{JCAP} {\bfseries
  07} (2012) 043} [\href{https://arxiv.org/abs/1205.0007}{{\ttfamily
  1205.0007}}].

\bibitem{Garcia-Cely:2016pse}
C.~Garcia-Cely and J.~Heeck, \emph{{Indirect searches of dark matter via
  polynomial spectral features}},
  \href{https://doi.org/10.1088/1475-7516/2016/08/023}{\emph{JCAP} {\bfseries
  08} (2016) 023} [\href{https://arxiv.org/abs/1605.08049}{{\ttfamily
  1605.08049}}].

\bibitem{Aghanim:2018eyx}
{\scshape Planck} collaboration, \emph{{Planck 2018 results. VI. Cosmological
  parameters}},
  \href{https://doi.org/10.1051/0004-6361/201833910}{\emph{Astron. Astrophys.}
  {\bfseries 641} (2020) A6}
  [\href{https://arxiv.org/abs/1807.06209}{{\ttfamily 1807.06209}}].

\bibitem{Hiroshima:2018kfv}
N.~Hiroshima, S.~Ando and T.~Ishiyama, \emph{{Modeling evolution of dark matter
  substructure and annihilation boost}},
  \href{https://doi.org/10.1103/PhysRevD.97.123002}{\emph{Phys. Rev. D}
  {\bfseries 97} (2018) 123002}
  [\href{https://arxiv.org/abs/1803.07691}{{\ttfamily 1803.07691}}].

\bibitem{Ando:2019xlm}
S.~Ando, T.~Ishiyama and N.~Hiroshima, \emph{{Halo Substructure Boosts to the
  Signatures of Dark Matter Annihilation}},
  \href{https://doi.org/10.3390/galaxies7030068}{\emph{Galaxies} {\bfseries 7}
  (2019) 68} [\href{https://arxiv.org/abs/1903.11427}{{\ttfamily 1903.11427}}].

\bibitem{Capozzi:2018ubv}
F.~Capozzi, E.~Lisi, A.~Marrone and A.~Palazzo, \emph{{Current unknowns in the
  three neutrino framework}},
  \href{https://doi.org/10.1016/j.ppnp.2018.05.005}{\emph{Prog. Part. Nucl.
  Phys.} {\bfseries 102} (2018) 48}
  [\href{https://arxiv.org/abs/1804.09678}{{\ttfamily 1804.09678}}].

\bibitem{Capozzi:2017ipn}
F.~Capozzi, E.~Di~Valentino, E.~Lisi, A.~Marrone, A.~Melchiorri and A.~Palazzo,
  \emph{{Global constraints on absolute neutrino masses and their ordering}},
  \href{https://doi.org/10.1103/PhysRevD.95.096014}{\emph{Phys. Rev. D}
  {\bfseries 95} (2017) 096014}
  [\href{https://arxiv.org/abs/2003.08511}{{\ttfamily 2003.08511}}].

\bibitem{Honda:2015fha}
M.~Honda, M.~Sajjad~Athar, T.~Kajita, K.~Kasahara and S.~Midorikawa,
  \emph{{Atmospheric neutrino flux calculation using the NRLMSISE-00
  atmospheric model}},
  \href{https://doi.org/10.1103/PhysRevD.92.023004}{\emph{Phys. Rev.}
  {\bfseries D92} (2015) 023004}
  [\href{https://arxiv.org/abs/1502.03916}{{\ttfamily 1502.03916}}].

\bibitem{Gorski_2005}
K.~M. Gorski, E.~Hivon, A.~J. Banday, B.~D. Wandelt, F.~K. Hansen, M.~Reinecke
  et~al., \emph{Healpix: A framework for high‐resolution discretization and
  fast analysis of data distributed on the sphere},
  \href{https://doi.org/10.1086/427976}{\emph{The Astrophysical Journal}
  {\bfseries 622} (2005) 759–771}.

\bibitem{Adrian-Martinez:2016fdl}
{\scshape KM3Net} collaboration, \emph{{Letter of intent for KM3NeT 2.0}},
  \href{https://doi.org/10.1088/0954-3899/43/8/084001}{\emph{J. Phys. G}
  {\bfseries 43} (2016) 084001}
  [\href{https://arxiv.org/abs/1601.07459}{{\ttfamily 1601.07459}}].

\bibitem{Gandhi:1998ri}
R.~Gandhi, C.~Quigg, M.~H. Reno and I.~Sarcevic, \emph{{Neutrino interactions
  at ultrahigh-energies}},
  \href{https://doi.org/10.1103/PhysRevD.58.093009}{\emph{Phys. Rev. D}
  {\bfseries 58} (1998) 093009}
  [\href{https://arxiv.org/abs/hep-ph/9807264}{{\ttfamily hep-ph/9807264}}].

\bibitem{Cowan_2011}
G.~Cowan, K.~Cranmer, E.~Gross and O.~Vitells, \emph{Asymptotic formulae for
  likelihood-based tests of new physics},
  \href{https://doi.org/10.1140/epjc/s10052-011-1554-0}{\emph{The European
  Physical Journal C} {\bfseries 71} (2011) }.

\bibitem{Griest:1989wd}
K.~Griest and M.~Kamionkowski, \emph{{Unitarity Limits on the Mass and Radius
  of Dark Matter Particles}},
  \href{https://doi.org/10.1103/PhysRevLett.64.615}{\emph{Phys. Rev. Lett.}
  {\bfseries 64} (1990) 615}.

\bibitem{Abercrombie:2015wmb}
D.~Abercrombie et~al., \emph{{Dark Matter Benchmark Models for Early LHC Run-2
  Searches: Report of the ATLAS/CMS Dark Matter Forum}},
  \href{https://doi.org/10.1016/j.dark.2019.100371}{\emph{Phys. Dark Univ.}
  {\bfseries 27} (2020) 100371}
  [\href{https://arxiv.org/abs/1507.00966}{{\ttfamily 1507.00966}}].

\bibitem{Boveia:2016mrp}
G.~Busoni et~al., \emph{{Recommendations on presenting LHC searches for missing
  transverse energy signals using simplified $s$-channel models of dark
  matter}}, \href{https://doi.org/10.1016/j.dark.2019.100365}{\emph{Phys. Dark
  Univ.} {\bfseries 27} (2020) 100365}
  [\href{https://arxiv.org/abs/1603.04156}{{\ttfamily 1603.04156}}].

\bibitem{Abdallah:2015ter}
J.~Abdallah et~al., \emph{{Simplified Models for Dark Matter Searches at the
  LHC}}, \href{https://doi.org/10.1016/j.dark.2015.08.001}{\emph{Phys. Dark
  Univ.} {\bfseries 9-10} (2015) 8}
  [\href{https://arxiv.org/abs/1506.03116}{{\ttfamily 1506.03116}}].

\bibitem{Kahlhoefer:2017dnp}
F.~Kahlhoefer, \emph{{Review of LHC Dark Matter Searches}},
  \href{https://doi.org/10.1142/S0217751X1730006X}{\emph{Int. J. Mod. Phys. A}
  {\bfseries 32} (2017) 1730006}
  [\href{https://arxiv.org/abs/1702.02430}{{\ttfamily 1702.02430}}].

\bibitem{Arcadi:2017kky}
G.~Arcadi, M.~Dutra, P.~Ghosh, M.~Lindner, Y.~Mambrini, M.~Pierre et~al.,
  \emph{{The waning of the WIMP? A review of models, searches, and
  constraints}},
  \href{https://doi.org/10.1140/epjc/s10052-018-5662-y}{\emph{Eur. Phys. J. C}
  {\bfseries 78} (2018) 203}
  [\href{https://arxiv.org/abs/1703.07364}{{\ttfamily 1703.07364}}].

\bibitem{Arina:2018zcq}
C.~Arina, \emph{{Impact of cosmological and astrophysical constraints on dark
  matter simplified models}},
  \href{https://doi.org/10.3389/fspas.2018.00030}{\emph{Front. Astron. Space
  Sci.} {\bfseries 5} (2018) 30}
  [\href{https://arxiv.org/abs/1805.04290}{{\ttfamily 1805.04290}}].

\bibitem{Blennow:2019fhy}
M.~Blennow, E.~Fernandez-Martinez, A.~Olivares-Del~Campo, S.~Pascoli,
  S.~Rosauro-Alcaraz and A.~Titov, \emph{{Neutrino Portals to Dark Matter}},
  \href{https://doi.org/10.1140/epjc/s10052-019-7060-5}{\emph{Eur. Phys. J. C}
  {\bfseries 79} (2019) 555}
  [\href{https://arxiv.org/abs/1903.00006}{{\ttfamily 1903.00006}}].

\bibitem{Alloul:2013bka}
A.~Alloul, N.~D. Christensen, C.~Degrande, C.~Duhr and B.~Fuks,
  \emph{{FeynRules 2.0 - A complete toolbox for tree-level phenomenology}},
  \href{https://doi.org/10.1016/j.cpc.2014.04.012}{\emph{Comput. Phys. Commun.}
  {\bfseries 185} (2014) 2250}
  [\href{https://arxiv.org/abs/1310.1921}{{\ttfamily 1310.1921}}].

\bibitem{Ambrogi:2018jqj}
F.~Ambrogi et~al., \emph{{MadDM v.3.0: a Comprehensive Tool for Dark Matter
  Studies}}, \href{https://doi.org/10.1016/j.dark.2018.11.009}{\emph{Phys. Dark
  Univ.} {\bfseries 24} (2019) 100249}
  [\href{https://arxiv.org/abs/1804.00044}{{\ttfamily 1804.00044}}].

\bibitem{Ando:2020yyk}
S.~Ando, A.~Geringer-Sameth, N.~Hiroshima, S.~Hoof, R.~Trotta and M.~G. Walker,
  \emph{{Structure formation models weaken limits on WIMP dark matter from
  dwarf spheroidal galaxies}},
  \href{https://doi.org/10.1103/PhysRevD.102.061302}{\emph{Phys. Rev. D}
  {\bfseries 102} (2020) 061302}
  [\href{https://arxiv.org/abs/2002.11956}{{\ttfamily 2002.11956}}].

\bibitem{Sjostrand:2014zea}
T.~Sjöstrand, S.~Ask, J.~R. Christiansen, R.~Corke, N.~Desai, P.~Ilten et~al.,
  \emph{{An Introduction to PYTHIA 8.2}},
  \href{https://doi.org/10.1016/j.cpc.2015.01.024}{\emph{Comput. Phys. Commun.}
  {\bfseries 191} (2015) 159}
  [\href{https://arxiv.org/abs/1410.3012}{{\ttfamily 1410.3012}}].

\bibitem{Queiroz:2016zwd}
F.~S. Queiroz, C.~E. Yaguna and C.~Weniger, \emph{{Gamma-ray Limits on Neutrino
  Lines}}, \href{https://doi.org/10.1088/1475-7516/2016/05/050}{\emph{JCAP}
  {\bfseries 05} (2016) 050}
  [\href{https://arxiv.org/abs/1602.05966}{{\ttfamily 1602.05966}}].

\bibitem{Blumenthal:1970gc}
G.~R. Blumenthal and R.~J. Gould, \emph{{Bremsstrahlung, synchrotron radiation,
  and compton scattering of high-energy electrons traversing dilute gases}},
  \href{https://doi.org/10.1103/RevModPhys.42.237}{\emph{Rev. Mod. Phys.}
  {\bfseries 42} (1970) 237}.

\bibitem{Jeltema:2008ax}
T.~E. Jeltema and S.~Profumo, \emph{{Searching for Dark Matter with X-ray
  Observations of Local Dwarf Galaxies}},
  \href{https://doi.org/10.1086/591495}{\emph{Astrophys. J.} {\bfseries 686}
  (2008) 1045} [\href{https://arxiv.org/abs/0805.1054}{{\ttfamily 0805.1054}}].

\bibitem{Natarajan:2015hma}
A.~Natarajan, J.~E. Aguirre, K.~Spekkens and B.~S. Mason, \emph{{Green Bank
  Telescope Constraints on Dark Matter Annihilation in Segue I}},
  \href{https://arxiv.org/abs/1507.03589}{{\ttfamily 1507.03589}}.

\bibitem{Kar:2019cqo}
A.~Kar, S.~Mitra, B.~Mukhopadhyaya and T.~R. Choudhury, \emph{{Heavy dark
  matter particle annihilation in dwarf spheroidal galaxies: radio signals at
  the SKA telescope}},
  \href{https://doi.org/10.1103/PhysRevD.101.023015}{\emph{Phys. Rev. D}
  {\bfseries 101} (2020) 023015}
  [\href{https://arxiv.org/abs/1905.11426}{{\ttfamily 1905.11426}}].

\bibitem{Buch:2015iya}
J.~Buch, M.~Cirelli, G.~Giesen and M.~Taoso, \emph{{PPPC 4 DM secondary: A Poor
  Particle Physicist Cookbook for secondary radiation from Dark Matter}},
  \href{https://doi.org/10.1088/1475-7516/2015/9/037}{\emph{JCAP} {\bfseries
  09} (2015) 037} [\href{https://arxiv.org/abs/1505.01049}{{\ttfamily
  1505.01049}}].

\bibitem{DEramo:2016gos}
F.~D'Eramo, B.~J. Kavanagh and P.~Panci, \emph{{You can hide but you have to
  run: direct detection with vector mediators}},
  \href{https://doi.org/10.1007/JHEP08(2016)111}{\emph{JHEP} {\bfseries 08}
  (2016) 111} [\href{https://arxiv.org/abs/1605.04917}{{\ttfamily
  1605.04917}}].

\bibitem{DEramo:2017zqw}
F.~D'Eramo, B.~J. Kavanagh and P.~Panci, \emph{{Probing Leptophilic Dark
  Sectors with Hadronic Processes}},
  \href{https://doi.org/10.1016/j.physletb.2017.05.063}{\emph{Phys. Lett. B}
  {\bfseries 771} (2017) 339}
  [\href{https://arxiv.org/abs/1702.00016}{{\ttfamily 1702.00016}}].

\bibitem{Cerdeno:2019vpd}
D.~Cerdeno, A.~Cheek, P.~Martin-Ramiro and J.~Moreno, \emph{{B anomalies and
  dark matter: a complex connection}},
  \href{https://doi.org/10.1140/epjc/s10052-019-6979-x}{\emph{Eur. Phys. J. C}
  {\bfseries 79} (2019) 517}
  [\href{https://arxiv.org/abs/1902.01789}{{\ttfamily 1902.01789}}].

\bibitem{Aprile:2018dbl}
{\scshape XENON} collaboration, \emph{{Dark Matter Search Results from a One
  Ton-Year Exposure of XENON1T}},
  \href{https://doi.org/10.1103/PhysRevLett.121.111302}{\emph{Phys. Rev. Lett.}
  {\bfseries 121} (2018) 111302}
  [\href{https://arxiv.org/abs/1805.12562}{{\ttfamily 1805.12562}}].

\bibitem{Cerdeno:2018bty}
D.~Cerdeno, A.~Cheek, E.~Reid and H.~Schulz, \emph{{Surrogate Models for Direct
  Dark Matter Detection}},
  \href{https://doi.org/10.1088/1475-7516/2018/08/011}{\emph{JCAP} {\bfseries
  08} (2018) 011} [\href{https://arxiv.org/abs/1802.03174}{{\ttfamily
  1802.03174}}].

\bibitem{Bozorgnia:2016ogo}
N.~Bozorgnia, F.~Calore, M.~Schaller, M.~Lovell, G.~Bertone, C.~S. Frenk
  et~al., \emph{{Simulated Milky Way analogues: implications for dark matter
  direct searches}},
  \href{https://doi.org/10.1088/1475-7516/2016/05/024}{\emph{JCAP} {\bfseries
  05} (2016) 024} [\href{https://arxiv.org/abs/1601.04707}{{\ttfamily
  1601.04707}}].

\bibitem{Ibarra:2018yxq}
A.~Ibarra, B.~J. Kavanagh and A.~Rappelt, \emph{{Bracketing the impact of
  astrophysical uncertainties on local dark matter searches}},
  \href{https://doi.org/10.1088/1475-7516/2018/12/018}{\emph{JCAP} {\bfseries
  12} (2018) 018} [\href{https://arxiv.org/abs/1806.08714}{{\ttfamily
  1806.08714}}].

\bibitem{Aad:2019qnd}
{\scshape ATLAS} collaboration, \emph{{Searches for electroweak production of
  supersymmetric particles with compressed mass spectra in $\sqrt{s}=$ 13 TeV
  $pp$ collisions with the ATLAS detector}},
  \href{https://doi.org/10.1103/PhysRevD.101.052005}{\emph{Phys. Rev. D}
  {\bfseries 101} (2020) 052005}
  [\href{https://arxiv.org/abs/1911.12606}{{\ttfamily 1911.12606}}].

\bibitem{Aad:2019fac}
{\scshape ATLAS} collaboration, \emph{{Search for high-mass dilepton resonances
  using 139 fb$^{-1}$ of $pp$ collision data collected at $\sqrt{s}=$13 TeV
  with the ATLAS detector}},
  \href{https://doi.org/10.1016/j.physletb.2019.07.016}{\emph{Phys. Lett. B}
  {\bfseries 796} (2019) 68}
  [\href{https://arxiv.org/abs/1903.06248}{{\ttfamily 1903.06248}}].

\bibitem{Aad:2019hjw}
{\scshape ATLAS} collaboration, \emph{{Search for new resonances in mass
  distributions of jet pairs using 139 fb$^{-1}$ of $pp$ collisions at
  $\sqrt{s}=13$ TeV with the ATLAS detector}},
  \href{https://doi.org/10.1007/JHEP03(2020)145}{\emph{JHEP} {\bfseries 03}
  (2020) 145} [\href{https://arxiv.org/abs/1910.08447}{{\ttfamily
  1910.08447}}].

\bibitem{Aaboud:2018fzt}
{\scshape ATLAS} collaboration, \emph{{Search for low-mass dijet resonances
  using trigger-level jets with the ATLAS detector in $pp$ collisions at
  $\sqrt{s}=13$ TeV}},
  \href{https://doi.org/10.1103/PhysRevLett.121.081801}{\emph{Phys. Rev. Lett.}
  {\bfseries 121} (2018) 081801}
  [\href{https://arxiv.org/abs/1804.03496}{{\ttfamily 1804.03496}}].

\bibitem{ATLAS:2020jhr}
{\scshape ATLAS} collaboration, \emph{{Search for dark matter in association
  with an energetic photon in pp collisions at s\ensuremath{\sqrt{}}=13 TeV
  with the ATLAS detector}}, .

\bibitem{ATLAS:2020wzf}
{\scshape ATLAS} collaboration, \emph{{Search for new phenomena in events with
  jets and missing transverse momentum in p p collisions at $\sqrt{s}$ = 13 TeV
  with the ATLAS detector}}, .

\bibitem{Zyla:2020zbs}
{\scshape Particle Data Group} collaboration, \emph{{Review of Particle
  Physics}}, \href{https://doi.org/10.1093/ptep/ptaa104}{\emph{PTEP} {\bfseries
  2020} (2020) 083C01}.

\bibitem{TheBABAR:2016rlg}
{\scshape BaBar} collaboration, \emph{{Search for a muonic dark force at
  BABAR}}, \href{https://doi.org/10.1103/PhysRevD.94.011102}{\emph{Phys. Rev.
  D} {\bfseries 94} (2016) 011102}
  [\href{https://arxiv.org/abs/1606.03501}{{\ttfamily 1606.03501}}].

\bibitem{Sirunyan:2018nnz}
{\scshape CMS} collaboration, \emph{{Search for an $L_{\mu}-L_{\tau}$ gauge
  boson using Z$\to4\mu$ events in proton-proton collisions at $\sqrt{s} =$ 13
  TeV}}, \href{https://doi.org/10.1016/j.physletb.2019.01.072}{\emph{Phys.
  Lett. B} {\bfseries 792} (2019) 345}
  [\href{https://arxiv.org/abs/1808.03684}{{\ttfamily 1808.03684}}].

\bibitem{Altmannshofer:2014pba}
W.~Altmannshofer, S.~Gori, M.~Pospelov and I.~Yavin, \emph{{Neutrino Trident
  Production: A Powerful Probe of New Physics with Neutrino Beams}},
  \href{https://doi.org/10.1103/PhysRevLett.113.091801}{\emph{Phys. Rev. Lett.}
  {\bfseries 113} (2014) 091801}
  [\href{https://arxiv.org/abs/1406.2332}{{\ttfamily 1406.2332}}].

\bibitem{Kaneta:2016uyt}
Y.~Kaneta and T.~Shimomura, \emph{{On the possibility of a search for the
  $L_\mu - L_\tau$ gauge boson at Belle-II and neutrino beam experiments}},
  \href{https://doi.org/10.1093/ptep/ptx050}{\emph{PTEP} {\bfseries 2017}
  (2017) 053B04} [\href{https://arxiv.org/abs/1701.00156}{{\ttfamily
  1701.00156}}].

\bibitem{Bauer:2018onh}
M.~Bauer, P.~Foldenauer and J.~Jaeckel, \emph{{Hunting All the Hidden
  Photons}}, \href{https://doi.org/10.1007/JHEP07(2018)094}{\emph{JHEP}
  {\bfseries 18} (2020) 094}
  [\href{https://arxiv.org/abs/1803.05466}{{\ttfamily 1803.05466}}].

\bibitem{Pospelov:2007mp}
M.~Pospelov, A.~Ritz and M.~B. Voloshin, \emph{{Secluded WIMP Dark Matter}},
  \href{https://doi.org/10.1016/j.physletb.2008.02.052}{\emph{Phys. Lett. B}
  {\bfseries 662} (2008) 53} [\href{https://arxiv.org/abs/0711.4866}{{\ttfamily
  0711.4866}}].

\bibitem{Arina:2020udz}
C.~Arina, B.~Fuks and L.~Mantani, \emph{{A universal framework for t-channel
  dark matter models}},
  \href{https://doi.org/10.1140/epjc/s10052-020-7933-7}{\emph{Eur. Phys. J. C}
  {\bfseries 80} (2020) 409}
  [\href{https://arxiv.org/abs/2001.05024}{{\ttfamily 2001.05024}}].

\bibitem{Arina:2020tuw}
C.~Arina, B.~Fuks, L.~Mantani, H.~Mies, L.~Panizzi and J.~Salko, \emph{{Closing
  in on $t$-channel simplified dark matter models}},
  \href{https://doi.org/10.1016/j.physletb.2020.136038}{\emph{Phys. Lett. B}
  {\bfseries 813} (2021) 136038}
  [\href{https://arxiv.org/abs/2010.07559}{{\ttfamily 2010.07559}}].

\bibitem{Ibarra:2015fqa}
A.~Ibarra and S.~Wild, \emph{{Dirac dark matter with a charged mediator: a
  comprehensive one-loop analysis of the direct detection phenomenology}},
  \href{https://doi.org/10.1088/1475-7516/2015/05/047}{\emph{JCAP} {\bfseries
  05} (2015) 047} [\href{https://arxiv.org/abs/1704.03850}{{\ttfamily
  1704.03850}}].

\bibitem{DAmbrosio:2002vsn}
G.~D'Ambrosio, G.~F. Giudice, G.~Isidori and A.~Strumia, \emph{{Minimal flavor
  violation: An Effective field theory approach}},
  \href{https://doi.org/10.1016/S0550-3213(02)00836-2}{\emph{Nucl. Phys. B}
  {\bfseries 645} (2002) 155}
  [\href{https://arxiv.org/abs/hep-ph/0207036}{{\ttfamily hep-ph/0207036}}].

\bibitem{Kopp:2014tsa}
J.~Kopp, L.~Michaels and J.~Smirnov, \emph{{Loopy Constraints on Leptophilic
  Dark Matter and Internal Bremsstrahlung}},
  \href{https://doi.org/10.1088/1475-7516/2014/04/022}{\emph{JCAP} {\bfseries
  04} (2014) 022} [\href{https://arxiv.org/abs/1401.6457}{{\ttfamily
  1401.6457}}].

\bibitem{Kavanagh:2018xeh}
B.~J. Kavanagh, P.~Panci and R.~Ziegler, \emph{{Faint Light from Dark Matter:
  Classifying and Constraining Dark Matter-Photon Effective Operators}},
  \href{https://doi.org/10.1007/JHEP04(2019)089}{\emph{JHEP} {\bfseries 04}
  (2019) 089} [\href{https://arxiv.org/abs/1810.00033}{{\ttfamily
  1810.00033}}].

\bibitem{Arina:2020mxo}
C.~Arina, A.~Cheek, K.~Mimasu and L.~Pagani, \emph{{Light and Darkness:
  consistently coupling dark matter to photons via effective operators}},
  \href{https://arxiv.org/abs/2005.12789}{{\ttfamily 2005.12789}}.

\bibitem{1931AnP...403..257S}
A.~{Sommerfeld}, \emph{{{\"U}ber die Beugung und Bremsung der Elektronen}},
  \href{https://doi.org/10.1002/andp.19314030302}{\emph{Annalen der Physik}
  {\bfseries 403} (1931) 257}.

\bibitem{Hambye:2005tk}
T.~Hambye, M.~Raidal and A.~Strumia, \emph{{Efficiency and maximal CP-asymmetry
  of scalar triplet leptogenesis}},
  \href{https://doi.org/10.1016/j.physletb.2005.11.007}{\emph{Phys. Lett. B}
  {\bfseries 632} (2006) 667}
  [\href{https://arxiv.org/abs/hep-ph/0510008}{{\ttfamily hep-ph/0510008}}].

\bibitem{Arina:2011cu}
C.~Arina and N.~Sahu, \emph{{Asymmetric Inelastic Inert Doublet Dark Matter
  from Triplet Scalar Leptogenesis}},
  \href{https://doi.org/10.1016/j.nuclphysb.2011.09.014}{\emph{Nucl. Phys. B}
  {\bfseries 854} (2012) 666}
  [\href{https://arxiv.org/abs/1108.3967}{{\ttfamily 1108.3967}}].

\bibitem{CentellesChulia:2020dfh}
S.~Centelles~Chuli\'a, R.~Srivastava and A.~Vicente, \emph{{The Inverse Seesaw
  Family: Dirac And Majorana}},
  \href{https://arxiv.org/abs/2011.06609}{{\ttfamily 2011.06609}}.

\bibitem{Andrea:2011ws}
J.~Andrea, B.~Fuks and F.~Maltoni, \emph{{Monotops at the LHC}},
  \href{https://doi.org/10.1103/PhysRevD.84.074025}{\emph{Phys. Rev. D}
  {\bfseries 84} (2011) 074025}
  [\href{https://arxiv.org/abs/1106.6199}{{\ttfamily 1106.6199}}].

\bibitem{Backovic:2015soa}
M.~Backovi\'c, M.~Kr\"amer, F.~Maltoni, A.~Martini, K.~Mawatari and M.~Pellen,
  \emph{{Higher-order QCD predictions for dark matter production at the LHC in
  simplified models with s-channel mediators}},
  \href{https://doi.org/10.1140/epjc/s10052-015-3700-6}{\emph{Eur. Phys. J. C}
  {\bfseries 75} (2015) 482}
  [\href{https://arxiv.org/abs/1508.05327}{{\ttfamily 1508.05327}}].

\bibitem{Mattelaer:2015haa}
O.~Mattelaer and E.~Vryonidou, \emph{{Dark matter production through
  loop-induced processes at the LHC: the s-channel mediator case}},
  \href{https://doi.org/10.1140/epjc/s10052-015-3665-5}{\emph{Eur. Phys. J. C}
  {\bfseries 75} (2015) 436}
  [\href{https://arxiv.org/abs/1508.00564}{{\ttfamily 1508.00564}}].

\bibitem{Neubert:2015fka}
M.~Neubert, J.~Wang and C.~Zhang, \emph{{Higher-Order QCD Predictions for Dark
  Matter Production in Mono-$Z$ Searches at the LHC}},
  \href{https://doi.org/10.1007/JHEP02(2016)082}{\emph{JHEP} {\bfseries 02}
  (2016) 082} [\href{https://arxiv.org/abs/1509.05785}{{\ttfamily
  1509.05785}}].

\bibitem{Aaboud:2019yqu}
{\scshape ATLAS} collaboration, \emph{{Constraints on mediator-based dark
  matter and scalar dark energy models using $\sqrt s = 13$ TeV $pp$ collision
  data collected by the ATLAS detector}},
  \href{https://doi.org/10.1007/JHEP05(2019)142}{\emph{JHEP} {\bfseries 05}
  (2019) 142} [\href{https://arxiv.org/abs/1903.01400}{{\ttfamily
  1903.01400}}].

\bibitem{Sirunyan:2018exx}
{\scshape CMS} collaboration, \emph{{Search for high-mass resonances in
  dilepton final states in proton-proton collisions at $\sqrt{s}=$ 13 TeV}},
  \href{https://doi.org/10.1007/JHEP06(2018)120}{\emph{JHEP} {\bfseries 06}
  (2018) 120} [\href{https://arxiv.org/abs/1803.06292}{{\ttfamily
  1803.06292}}].

\bibitem{Stueckelberg:1900zz}
E.~Stueckelberg, \emph{{Interaction energy in electrodynamics and in the field
  theory of nuclear forces}},
  \href{https://doi.org/10.5169/seals-110852}{\emph{Helv. Phys. Acta}
  {\bfseries 11} (1938) 225}.

\bibitem{Kors:2005uz}
B.~Kors and P.~Nath, \emph{{Aspects of the Stueckelberg extension}},
  \href{https://doi.org/10.1088/1126-6708/2005/07/069}{\emph{JHEP} {\bfseries
  07} (2005) 069} [\href{https://arxiv.org/abs/hep-ph/0503208}{{\ttfamily
  hep-ph/0503208}}].

\bibitem{Basso:2011na}
L.~Basso, S.~Moretti and G.~M. Pruna, \emph{{Theoretical constraints on the
  couplings of non-exotic minimal $Z'$ bosons}},
  \href{https://doi.org/10.1007/JHEP08(2011)122}{\emph{JHEP} {\bfseries 08}
  (2011) 122} [\href{https://arxiv.org/abs/1106.4762}{{\ttfamily 1106.4762}}].

\bibitem{Carena:2004xs}
M.~Carena, A.~Daleo, B.~A. Dobrescu and T.~M.~P. Tait, \emph{{$Z^\prime$ gauge
  bosons at the Tevatron}},
  \href{https://doi.org/10.1103/PhysRevD.70.093009}{\emph{Phys. Rev. D}
  {\bfseries 70} (2004) 093009}
  [\href{https://arxiv.org/abs/hep-ph/0408098}{{\ttfamily hep-ph/0408098}}].

\bibitem{Baek_2009}
S.~Baek and P.~Ko, \emph{{Phenomenology of U(1)(L(mu)-L(tau)) charged dark
  matter at PAMELA and colliders}},
  \href{https://doi.org/10.1088/1475-7516/2009/10/011}{\emph{JCAP} {\bfseries
  10} (2009) 011} [\href{https://arxiv.org/abs/0811.1646}{{\ttfamily
  0811.1646}}].

\bibitem{Biswas:2016yjr}
A.~Biswas, S.~Choubey and S.~Khan, \emph{{FIMP and Muon ($g-2$) in a
  U$(1)_{L_{\mu}-L_{\tau}}$ Model}},
  \href{https://doi.org/10.1007/JHEP02(2017)123}{\emph{JHEP} {\bfseries 02}
  (2017) 123} [\href{https://arxiv.org/abs/1612.03067}{{\ttfamily
  1612.03067}}].

\bibitem{Altmannshofer:2016jzy}
W.~Altmannshofer, S.~Gori, S.~Profumo and F.~S. Queiroz, \emph{{Explaining dark
  matter and B decay anomalies with an $L_\mu - L_\tau$ model}},
  \href{https://doi.org/10.1007/JHEP12(2016)106}{\emph{JHEP} {\bfseries 12}
  (2016) 106} [\href{https://arxiv.org/abs/1609.04026}{{\ttfamily
  1609.04026}}].

\bibitem{Arcadi:2018tly}
G.~Arcadi, T.~Hugle and F.~S. Queiroz, \emph{{The Dark $L_\mu - L_\tau$ Rises
  via Kinetic Mixing}},
  \href{https://doi.org/10.1016/j.physletb.2018.07.028}{\emph{Phys. Lett. B}
  {\bfseries 784} (2018) 151}
  [\href{https://arxiv.org/abs/1803.05723}{{\ttfamily 1803.05723}}].

\bibitem{Araki:2017wyg}
T.~Araki, S.~Hoshino, T.~Ota, J.~Sato and T.~Shimomura, \emph{{Detecting the
  $L_{\mu}-L_{\tau}$ gauge boson at Belle II}},
  \href{https://doi.org/10.1103/PhysRevD.95.055006}{\emph{Phys. Rev. D}
  {\bfseries 95} (2017) 055006}
  [\href{https://arxiv.org/abs/1702.01497}{{\ttfamily 1702.01497}}].

\bibitem{Aaij:2014ora}
{\scshape LHCb} collaboration, \emph{{Test of lepton universality using
  $B^{+}\rightarrow K^{+}\ell^{+}\ell^{-}$ decays}},
  \href{https://doi.org/10.1103/PhysRevLett.113.151601}{\emph{Phys. Rev. Lett.}
  {\bfseries 113} (2014) 151601}
  [\href{https://arxiv.org/abs/1406.6482}{{\ttfamily 1406.6482}}].

\bibitem{Aaij:2017vbb}
{\scshape LHCb} collaboration, \emph{{Test of lepton universality with $B^{0}
  \rightarrow K^{*0}\ell^{+}\ell^{-}$ decays}},
  \href{https://doi.org/10.1007/JHEP08(2017)055}{\emph{JHEP} {\bfseries 08}
  (2017) 055} [\href{https://arxiv.org/abs/1705.05802}{{\ttfamily
  1705.05802}}].

\bibitem{Aaij:2019wad}
{\scshape LHCb} collaboration, \emph{{Search for lepton-universality violation
  in $B^+\to K^+\ell^+\ell^-$ decays}},
  \href{https://doi.org/10.1103/PhysRevLett.122.191801}{\emph{Phys. Rev. Lett.}
  {\bfseries 122} (2019) 191801}
  [\href{https://arxiv.org/abs/1903.09252}{{\ttfamily 1903.09252}}].

\bibitem{Alguero:2019ptt}
M.~Alguer\'o, B.~Capdevila, A.~Crivellin, S.~Descotes-Genon, P.~Masjuan,
  J.~Matias et~al., \emph{{Emerging patterns of New Physics with and without
  Lepton Flavour Universal contributions}},
  \href{https://doi.org/10.1140/epjc/s10052-019-7216-3}{\emph{Eur. Phys. J. C}
  {\bfseries 79} (2019) 714}
  [\href{https://arxiv.org/abs/1903.09578}{{\ttfamily 1903.09578}}].

\bibitem{Altmannshofer:2019xda}
W.~Altmannshofer, J.~Davighi and M.~Nardecchia, \emph{{Gauging the accidental
  symmetries of the standard model, and implications for the flavor
  anomalies}}, \href{https://doi.org/10.1103/PhysRevD.101.015004}{\emph{Phys.
  Rev. D} {\bfseries 101} (2020) 015004}
  [\href{https://arxiv.org/abs/1909.02021}{{\ttfamily 1909.02021}}].

\bibitem{DiLuzio:2019jyq}
L.~Di~Luzio, M.~Kirk, A.~Lenz and T.~Rauh, \emph{{$\Delta M_s$ theory precision
  confronts flavour anomalies}},
  \href{https://doi.org/10.1007/JHEP12(2019)009}{\emph{JHEP} {\bfseries 12}
  (2019) 009} [\href{https://arxiv.org/abs/1909.11087}{{\ttfamily
  1909.11087}}].

\bibitem{Altmannshofer:2015mqa}
W.~Altmannshofer and I.~Yavin, \emph{{Predictions for lepton flavor
  universality violation in rare B decays in models with gauged $L_\mu -
  L_\tau$}}, \href{https://doi.org/10.1103/PhysRevD.92.075022}{\emph{Phys. Rev.
  D} {\bfseries 92} (2015) 075022}
  [\href{https://arxiv.org/abs/1508.07009}{{\ttfamily 1508.07009}}].

\bibitem{Bringmann:2016din}
T.~Bringmann, F.~Kahlhoefer, K.~Schmidt-Hoberg and P.~Walia, \emph{{Strong
  constraints on self-interacting dark matter with light mediators}},
  \href{https://doi.org/10.1103/PhysRevLett.118.141802}{\emph{Phys. Rev. Lett.}
  {\bfseries 118} (2017) 141802}
  [\href{https://arxiv.org/abs/1612.00845}{{\ttfamily 1612.00845}}].

\bibitem{Slatyer:2015jla}
T.~R. Slatyer, \emph{{Indirect dark matter signatures in the cosmic dark ages.
  I. Generalizing the bound on s-wave dark matter annihilation from Planck
  results}}, \href{https://doi.org/10.1103/PhysRevD.93.023527}{\emph{Phys. Rev.
  D} {\bfseries 93} (2016) 023527}
  [\href{https://arxiv.org/abs/1506.03811}{{\ttfamily 1506.03811}}].

\bibitem{ArkaniHamed:2008qn}
N.~Arkani-Hamed, D.~P. Finkbeiner, T.~R. Slatyer and N.~Weiner, \emph{{A Theory
  of Dark Matter}},
  \href{https://doi.org/10.1103/PhysRevD.79.015014}{\emph{Phys. Rev. D}
  {\bfseries 79} (2009) 015014}
  [\href{https://arxiv.org/abs/0810.0713}{{\ttfamily 0810.0713}}].

\bibitem{Bernal:2016gxb}
J.~L. Bernal, L.~Verde and A.~G. Riess, \emph{{The trouble with $H_0$}},
  \href{https://doi.org/10.1088/1475-7516/2016/10/019}{\emph{JCAP} {\bfseries
  10} (2016) 019} [\href{https://arxiv.org/abs/1607.05617}{{\ttfamily
  1607.05617}}].

\bibitem{Escudero:2019gzq}
M.~Escudero, D.~Hooper, G.~Krnjaic and M.~Pierre, \emph{{Cosmology with A Very
  Light L$_{\mu}$ \ensuremath{-} L$_{\tau}$ Gauge Boson}},
  \href{https://doi.org/10.1007/JHEP03(2019)071}{\emph{JHEP} {\bfseries 03}
  (2019) 071} [\href{https://arxiv.org/abs/1901.02010}{{\ttfamily
  1901.02010}}].

\bibitem{Bennett:2002jb}
{\scshape Muon g-2} collaboration, \emph{{Measurement of the positive muon
  anomalous magnetic moment to 0.7 ppm}},
  \href{https://doi.org/10.1103/PhysRevLett.89.101804}{\emph{Phys. Rev. Lett.}
  {\bfseries 89} (2002) 101804}
  [\href{https://arxiv.org/abs/hep-ex/0208001}{{\ttfamily hep-ex/0208001}}].

\bibitem{Bennett:2004pv}
{\scshape Muon g-2} collaboration, \emph{{Measurement of the negative muon
  anomalous magnetic moment to 0.7 ppm}},
  \href{https://doi.org/10.1103/PhysRevLett.92.161802}{\emph{Phys. Rev. Lett.}
  {\bfseries 92} (2004) 161802}
  [\href{https://arxiv.org/abs/hep-ex/0401008}{{\ttfamily hep-ex/0401008}}].

\bibitem{Bennett:2006fi}
{\scshape Muon g-2} collaboration, \emph{{Final Report of the Muon E821
  Anomalous Magnetic Moment Measurement at BNL}},
  \href{https://doi.org/10.1103/PhysRevD.73.072003}{\emph{Phys. Rev. D}
  {\bfseries 73} (2006) 072003}
  [\href{https://arxiv.org/abs/hep-ex/0602035}{{\ttfamily hep-ex/0602035}}].

\bibitem{Roberts:2010cj}
B.~L. Roberts, \emph{{Status of the Fermilab Muon $(g-2)$ Experiment}},
  \href{https://doi.org/10.1088/1674-1137/34/6/021}{\emph{Chin. Phys. C}
  {\bfseries 34} (2010) 741} [\href{https://arxiv.org/abs/1001.2898}{{\ttfamily
  1001.2898}}].

\bibitem{Amaral:2020tga}
D.~W. P.~d. Amaral, D.~G. Cerdeno, P.~Foldenauer and E.~Reid, \emph{{Solar
  neutrino probes of the muon anomalous magnetic moment in the gauged $
  \mathrm{U}{(1)}_{L_{\mu }-{L}_{\tau }} $}},
  \href{https://doi.org/10.1007/JHEP12(2020)155}{\emph{JHEP} {\bfseries 20}
  (2020) 155} [\href{https://arxiv.org/abs/2006.11225}{{\ttfamily
  2006.11225}}].

\bibitem{Abe:2020sbr}
{\scshape Super-Kamiokande} collaboration, \emph{{Indirect search for dark
  matter from the Galactic Center and halo with the Super-Kamiokande
  detector}}, \href{https://doi.org/10.1103/PhysRevD.102.072002}{\emph{Phys.
  Rev. D} {\bfseries 102} (2020) 072002}
  [\href{https://arxiv.org/abs/2005.05109}{{\ttfamily 2005.05109}}].

\bibitem{Klop:2018ltd}
N.~Klop and S.~Ando, \emph{{Constraints on MeV dark matter using neutrino
  detectors and their implication for the 21-cm results}},
  \href{https://doi.org/10.1103/PhysRevD.98.103004}{\emph{Phys. Rev. D}
  {\bfseries 98} (2018) 103004}
  [\href{https://arxiv.org/abs/1809.00671}{{\ttfamily 1809.00671}}].

\bibitem{Asai:2020qlp}
K.~Asai, S.~Okawa and K.~Tsumura, \emph{{Search for U(1)$_{L_\mu-L_\tau}$
  charged Dark Matter with neutrino telescope}},
  \href{https://arxiv.org/abs/2011.03165}{{\ttfamily 2011.03165}}.

\bibitem{Ibarra:2013eda}
A.~Ibarra, H.~M. Lee, S.~L\'opez~Gehler, W.-I. Park and M.~Pato,
  \emph{{Gamma-ray boxes from axion-mediated dark matter}},
  \href{https://doi.org/10.1088/1475-7516/2013/05/016}{\emph{JCAP} {\bfseries
  05} (2013) 016} [\href{https://arxiv.org/abs/1303.6632}{{\ttfamily
  1303.6632}}].

\bibitem{Ibarra:2015tya}
A.~Ibarra, A.~S. Lamperstorfer, S.~L\'opez-Gehler, M.~Pato and G.~Bertone,
  \emph{{On the sensitivity of CTA to gamma-ray boxes from multi-TeV dark
  matter}}, \href{https://doi.org/10.1088/1475-7516/2016/06/E02,
  10.1088/1475-7516/2015/09/048}{\emph{JCAP} {\bfseries 1509} (2015) 048}
  [\href{https://arxiv.org/abs/1503.06797}{{\ttfamily 1503.06797}}].

\bibitem{Leane:2017vag}
R.~K. Leane, K.~C.~Y. Ng and J.~F. Beacom, \emph{{Powerful Solar Signatures of
  Long-Lived Dark Mediators}},
  \href{https://doi.org/10.1103/PhysRevD.95.123016}{\emph{Phys. Rev. D}
  {\bfseries 95} (2017) 123016}
  [\href{https://arxiv.org/abs/1703.04629}{{\ttfamily 1703.04629}}].

\bibitem{Arina:2017sng}
C.~Arina, M.~Backovi\'c, J.~Heisig and M.~Lucente, \emph{{Solar $\gamma$ rays
  as a complementary probe of dark matter}},
  \href{https://doi.org/10.1103/PhysRevD.96.063010}{\emph{Phys. Rev. D}
  {\bfseries 96} (2017) 063010}
  [\href{https://arxiv.org/abs/1703.08087}{{\ttfamily 1703.08087}}].

\bibitem{Hisano:2004ds}
J.~Hisano, S.~Matsumoto, M.~M. Nojiri and O.~Saito, \emph{{Non-perturbative
  effect on dark matter annihilation and gamma ray signature from galactic
  center}}, \href{https://doi.org/10.1103/PhysRevD.71.063528}{\emph{Phys. Rev.
  D} {\bfseries 71} (2005) 063528}
  [\href{https://arxiv.org/abs/hep-ph/0412403}{{\ttfamily hep-ph/0412403}}].

\bibitem{Iengo:2009ni}
R.~Iengo, \emph{{Sommerfeld enhancement: General results from field theory
  diagrams}}, \href{https://doi.org/10.1088/1126-6708/2009/05/024}{\emph{JHEP}
  {\bfseries 05} (2009) 024} [\href{https://arxiv.org/abs/0902.0688}{{\ttfamily
  0902.0688}}].

\bibitem{Cassel:2009wt}
S.~Cassel, \emph{{Sommerfeld factor for arbitrary partial wave processes}},
  \href{https://doi.org/10.1088/0954-3899/37/10/105009}{\emph{J. Phys.}
  {\bfseries G37} (2010) 105009}
  [\href{https://arxiv.org/abs/0903.5307}{{\ttfamily 0903.5307}}].

\bibitem{Arina:2010wv}
C.~Arina, F.-X. Josse-Michaux and N.~Sahu, \emph{{Constraining Sommerfeld
  Enhanced Annihilation Cross-sections of Dark Matter via Direct Searches}},
  \href{https://doi.org/10.1016/j.physletb.2010.06.037}{\emph{Phys. Lett. B}
  {\bfseries 691} (2010) 219}
  [\href{https://arxiv.org/abs/1004.0645}{{\ttfamily 1004.0645}}].

\bibitem{Riess:2016jrr}
A.~G. Riess et~al., \emph{{A 2.4\% Determination of the Local Value of the
  Hubble Constant}},
  \href{https://doi.org/10.3847/0004-637X/826/1/56}{\emph{Astrophys. J.}
  {\bfseries 826} (2016) 56}
  [\href{https://arxiv.org/abs/1604.01424}{{\ttfamily 1604.01424}}].

\bibitem{Riess:2018byc}
A.~G. Riess et~al., \emph{{Milky Way Cepheid Standards for Measuring Cosmic
  Distances and Application to Gaia DR2: Implications for the Hubble
  Constant}}, \href{https://doi.org/10.3847/1538-4357/aac82e}{\emph{Astrophys.
  J.} {\bfseries 861} (2018) 126}
  [\href{https://arxiv.org/abs/1804.10655}{{\ttfamily 1804.10655}}].

\bibitem{Riess:2019cxk}
A.~G. Riess, S.~Casertano, W.~Yuan, L.~M. Macri and D.~Scolnic, \emph{{Large
  Magellanic Cloud Cepheid Standards Provide a 1\% Foundation for the
  Determination of the Hubble Constant and Stronger Evidence for Physics beyond
  $\Lambda$CDM}},
  \href{https://doi.org/10.3847/1538-4357/ab1422}{\emph{Astrophys. J.}
  {\bfseries 876} (2019) 85}
  [\href{https://arxiv.org/abs/1903.07603}{{\ttfamily 1903.07603}}].

\end{thebibliography}\endgroup
\end{document}